\documentclass{aastex63}

\usepackage[utf8]{inputenc}
\usepackage{kotex}
\usepackage{kotex-logo}

\usepackage{color}
\usepackage{multirow}

\shorttitle{EW Boo}
\shortauthors{Kim et al.}

\graphicspath{{./}{figures/}}

\begin{document}

\title{A Photometric and Spectroscopic Study of the Short-Period Algol EW Bo\"{o}tis with a $\delta$ Sct Pulsator}

\correspondingauthor{Chun-Hwey Kim}
\email{kimch@chungbuk.ac.kr}

\author[0000-0002-6687-6318]{Hye-Young Kim}
\affiliation{Department of Astronomy and Space Science, Chungbuk National University, Cheongju 28644, Korea}

\author[0000-0002-8692-2588]{Kyeongsoo Hong}
\affiliation{Institute for Astrophysics, Chungbuk National University, Cheongju 28644, Republic of Korea}

\author[0000-0001-8591-4562]{Chun-Hwey Kim}
\affiliation{Department of Astronomy and Space Science, Chungbuk National University, Cheongju 28644, Korea}

\author[0000-0002-5739-9804]{Jae Woo Lee}
\affiliation{Korea Astronomy and Space Science Institute, Daejeon 34055, Korea}

\author[0000-0002-8394-7237]{Min-Ji Jeong}
\affiliation{Department of Astronomy and Space Science, Chungbuk National University, Cheongju 28644, Korea}

\author[0000-0001-9339-4456]{Jang-Ho Park}
\affiliation{Korea Astronomy and Space Science Institute, Daejeon 34055, Korea}

\author[0000-0003-2772-7528]{Mi-Hwa Song}
\affiliation{Department of Astronomy and Space Science, Chungbuk National University, Cheongju 28644, Korea}




\begin{abstract}

In this paper, we present TESS photometry and high-resolution spectra of the short-period Algol EW Boo. We obtained double-lined radial velocities (RVs) from the time-series spectra and measured the effective temperature of the primary star as $T_{\rm{eff,1}}$ = 8560 $\pm$ 118 K. For the orbital period study, we collected all times of minima available for over the last 30 years. It was found that the eclipse timing variation of the system could be represented by a periodic oscillation of 17.6 $\pm$ 0.3 years with a semi-amplitude of 0.0041 $\pm$ 0.0001 d. The orbital and physical parameters were derived by simultaneously analyzing the TESS light and RV curves using the Wilson--Devinney (WD) binary star modeling code. The component masses and radii were showed over 3\% precision: $M_{1}$ = 2.67 $\pm $ 0.08 M$_{\odot}$, $M_{2}$ = 0.43 $\pm $ 0.01 M$_{\odot}$, $R_{1}$ = 2.01 $\pm $ 0.02 R$_{\odot}$, and $R_{2}$ = 1.35 $\pm $ 0.01 R$_{\odot}$. Furthermore, multiple frequency analyses were performed for the light-curve residuals from the WD model. As a result, we detected 17 pressure-mode pulsations in the region of 40.15 $–$ 52.37 d$^{-1}$. The absolute dimensions and pulsation characteristics showed that the $\delta$ Sct pulsator was the more massive and hotter primary star of the EW Boo.

\end{abstract}


\section{Introduction} \label{sec:1}
Eclipsing binaries (EBs) are used to study dynamical interactions between stars, such as mass transfer or tidal distortion, which has an influence on their evolution. Their light and double-lined RVs curves provide fundamental data for both components, such as mass, radius, and luminosity \citep{Torres+2010}. Furthermore, the changes in the orbital period of EBs provide important clues for understanding the additional companions in systems, mass transfer from one component to another, mass loss from one (or both) component(s), and the magnetic activities of stars \citep{Kreine+2001aocd.book.....K} in astrophysics. In some cases, pulsation occurs in the components of the binary systems. Pulsating EBs can be a promising target for asteroseismic studies as well as understanding the effect of interactions between binary components on pulsations.

The study of stellar oscillations, known as asteroseismology, allows us to probe the internal structure of stars. $\delta$ Sct stars are intermediate-mass objects with a typical mass range of 1.5–2.5 M$_{\odot}$ \citep{Aerts+2010aste.book.....A}, which corresponds to the transition mass region at which the internal structure is inverted from low-mass stars (M $\lesssim$ 1.5 M$_{\odot}$) with radiative cores and convective envelopes to intermediate-mass stars (M $\gtrsim$ 2.5 M$_{\odot}$) with convective cores and radiative envelopes. $\delta$ Sct stars are multiperiodic pulsators that pulsate in low-order radial and non-radial pressure modes with a frequency range of 4–50 d$^{-1}$ \citep{Breger_2000ASPC..210....3B} owing to the $\kappa$ mechanism operating in the He II partial ionization zone \citep{Pamyatnykh_2000ASPC..210..215P}. Generally, their temperature corresponds to the A-F spectral types that lie on a classical instability strip \citep{Rodriguez&Breger_2001A&A...366..178R}. The red edge of the $\delta$ Sct instability strip overlaps with the blue edge of $\gamma$ Dor. The stars located in this intersectional region of the HR diagram show $p$ and $g$ mode pulsations simultaneously. However, it is necessary to determine the exact physical properties of the $\delta$ Sct stars to understand their asteroseismic characteristics and interior structures.

Recently, various space missions have contributed to in-depth studies on pulsating EBs \citep{Lampens_2021Galax...9...28L}. Among them, the Transiting Exoplanet Survey Satellite (TESS), launched on April 18, 2018, aimed to search for small exoplanets \citep{Ricker+2015JATIS...1a4003R}. Photometric transit events were monitored by surveying over 85$\%$ of the sky. The observations were conducted on the sky of the northern and southern hemispheres, which were divided into 13 sectors using four cameras covering the regions of $24^{\circ} \times 96^{\circ}$. Considering the space observations were not interrupted during the daytime, the TESS provided nearly continuous data of approximately 27.4 d. The acquisition of these dense light curves with high precision was highly beneficial for analyzing stellar eclipse events and detecting many pulsation frequencies in mmag units.

The eclipsing binary EW Boo was discovered by the $Hipparcos$ satellite \citep{ESA_1997yCat.1239....0E} as an Algol-type binary with a spectral type of A0. \citet{Soydugan+2006MNRAS.370.2013S} suggested it to be a candidate of the oscillating Algol type (oEA) binary. The first ground-based photometric observation of the system was made using the $V$ passband by \citet{Soydugan+2008CoAst.157..379S}. Through data analysis, they presented a surface temperature of 8179 K for the primary component and a mass ratio of 0.217 for the system. Additionally, it was suggested that the primary component displayed $\delta$ Sct-type oscillation with a period of $\sim$ 30 m ($f$ = $\sim$ 48 d$^{-1}$) and an amplitude of approximately 0.02 mag. However, no further information has been published on their solution parameters. Furthermore, some physical parameters of EW Boo were characterized by \citet{Soydugan&Kacar_2013AJ....145...87S} as $M_{1}$ = 1.4 M$_{\odot}$, $M_{2}$ = 0.31 M$_{\odot}$, $a$ = 4.7 $R_{\odot}$, and $R_{1}$ = 1.7 $R_{\odot}$. Recently, more detailed physical properties of EW Boo were independently studied by \citet[][hereinafter ZLW]{Zhang+2015AJ....149...96Z}, based solely on $BV$ photometry, and \citet[][hereinafter DG]{Dogruel+2015NewA...40...20D}, based on both $BVRI$ photometry and spectroscopy. Consistent with their work, EW Boo can be defined as a semi-detached binary star with a low-mass secondary star filling the Roche lobe, and a primary star as a pulsating star. However, there are non-negligible differences between the physical parameters determined by ZLW and DG (Compare ZLW’s Table 3 with DG’s Table 7). For example, ZLW and DG adopted the temperature of the primary star ($T_{1}$) at 7840 K and 8970 K, respectively, followed by obtaining significantly different temperature differences ($\Delta T = T_{1}-T_{2}$) between the primary and secondary stars as $\Delta T_{ZLW}$ = 3325 K and $\Delta T_{DG}$ = 3928 K, respectively. For orbital inclination, ZLW was determined to be 76$^{\circ}$.50 ($\pm$ 0.13), whereas the DG derived 74$^{\circ}$.279 ($\pm$ 0.016). Furthermore, the absolute dimensions obtained by ZLW and DG were quite different from each other: The absolute dimensions of the primary and secondary components were determined by ZLW as $M_{1}$ = 1.80 M$_{\odot}$, $M_{2}$ = 0.23 M$_{\odot}$, $R_{1}$ = 1.84 R$_{\odot}$, $R_{2}$ = 1.11 R$_{\odot}$, $L_{1}$ = 10.9 L$_{\odot}$ and $L_{1}$ = 0.4 L$_{\odot}$, whereas DG determined $M_{1}$ = 2.00 M$_{\odot}$, $M_{2}$ = 0.327 M$_{\odot}$, $R_{1}$ = 1.88 R$_{\odot}$, $R_{2}$ = 1.23 R$_{\odot}$, $L_{1}$ = 20.46 L$_{\odot}$ and $L_{1}$ = 0.88 L$_{\odot}$. Moreover, ZLW and DG exhibited different interpretations of the eclipse timing diagrams that were created using almost the same timing data. ZLW suggested a secular increase in the orbital period of approximately $+2.9 \times 10^{-9}$ d yr$^{-1}$, whereas DG recorded no such changes. Regarding the pulsation occurring in the primary component, ZLW found four pulsation frequencies ($f_{1}$ = 52.372 $\pm$ 0.001 d$^{-1}$, $f_{2}$ = 47.320 $\pm$ 0.001 d$^{-1}$, $f_{3}$ = 49.481 $\pm$ 0.002 d$^{-1}$, and $f_{4}$ = 50.296 $\pm$ 0.002 d$^{-1}$) in the $BV$ passband data, whereas DG found only two frequencies ($f_{1}$ = 49.4496 $\pm$ 0.0001 d$^{-1}$ and $f_{2}$ = 43.8991 $\pm$ 0.0002 d$^{-1}$) in the $V$ data. Using the Frequency Analysis and Mode Identification for Asteroseismology program \citep{Zima_2008CoAst.155...17Z}, ZLW suggested that the obtained frequencies can be attributed to pulsation in non-radial modes with an angular quantum number of $l$ = 1. On the contrary, using the model of \citet{Fitch_1981ApJ...249..218F}, DG presented the following two possible causes for $f_{1}$ and $f_{2}$: 1) either the fifth or fourth overtones of radial pulsation ($l$ = 0) or 2) the fifth and fourth-degree p-modes of non-radial pulsation ($l$ $>$ 1), respectively. Therefore, it is still unclear whether the pulsation of EW Boo is caused by radial or non-radial pulsations.

As mentioned above, looking at the short research history of the semi-detached binary star EW Boo with $\delta$ Sct-type pulsation, further research is needed in that the physical parameters of the star and the detailed properties of the pulsation are not unified among investigators. In this study, we carried out a photometric and spectroscopic observation project to determine the system parameters of pulsating eclipsing binaries and understand their pulsation properties \citep[for example,][]{Hong+2017AJ....153..247H, Lee+2018_OODra, Park+2020AJ....160..247P, Hong+2021AJ....161..137H}, which includes EW Boo. We presented the binary and pulsation properties of EW Boo based on precise time-series TESS photometric and our high-resolution spectroscopic observations. Section 2 presents the observations and data analysis. Sections 3 and 4 report the results of the period study and light curve synthesis. Section 5 describes the multi-frequency analysis and pulsation properties. Finally, Section 6 provides a summary of the results.

\section{Observation and Data Analysis} \label{sec:2}
\subsection{TESS photometry} \label{sec:2.1}
\begin{figure}[]
\plotone{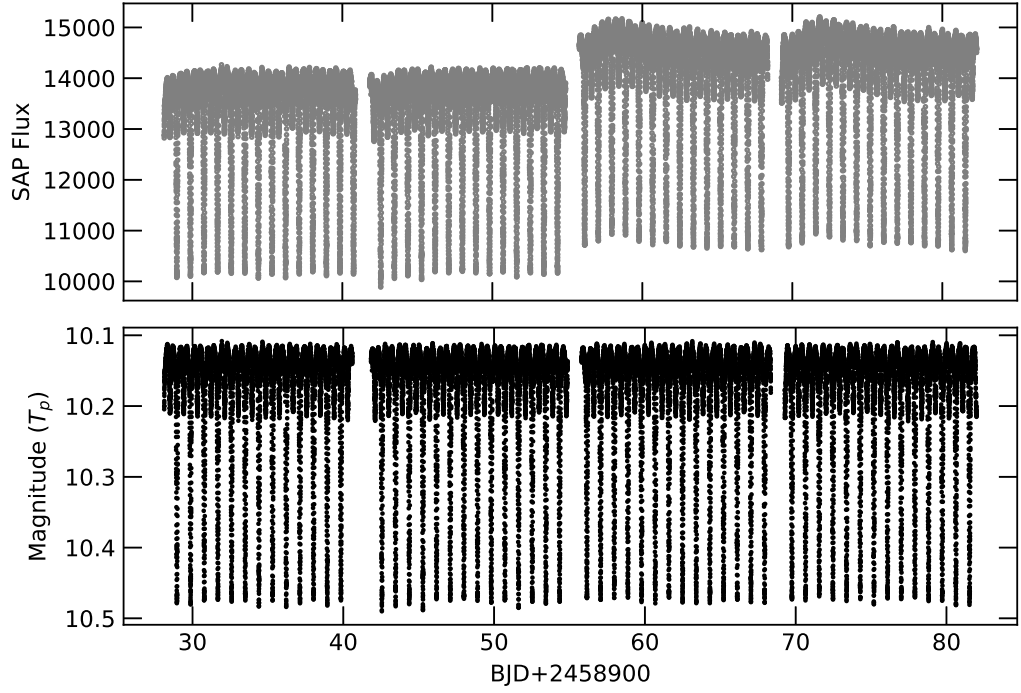}
\caption{The TESS photometric data of Sector 23 and 24 for EW Boo. The gray circles in the upper panel indicate the original SAP flux. In the bottom panel, the black circles represent the SAP magnitudes after detrending.  \label{fig:tess_detrending}}
\end{figure}

High-precision photometric observations of EW Boo were obtained by the TESS mission in a 2-min cadence mode during Sectors 23 and 24. The data was then observed from March 18 to April 16, 2020 (BJD 2,458,928.10 - 2,458,954.88) and April 16 to May 13, 2020 (BJD 2,458,955.79 – 2,458,982.29) using cameras 2 and 1, respectively. We used simple aperture photometry (SAP) flux data obtained from the Mikulski Archive for Space Telescopes (MAST) Portal\footnote{https://mast.stsci.edu/portal/Mashup/Clients/Mast/Portal.html}. The trends in the raw SAP data were eliminated by applying a least-squares spline fit to the outside ellipses of each sector. Through visual inspection, we eliminated the light curve outliers that occurred during this process, and consequently, a total of 36,550 TESS measurements were acquired. Then, the TESS magnitude of $T_{p}$ = +10.133 mag \citep{Stassun+2019AJ....158..138S} was applied to convert the SAP fluxes and their errors to a magnitude scale. The mean error of the SAP magnitude was 0.133 mmag. Figure~\ref{fig:tess_detrending} shows the TESS light curves of EW Boo before and after detrending.

\subsection{Spectroscopy} \label{sec:2.2}
We observed a total of 39 high-resolution spectra from the Bohyunsan Observatory Echelle Spectrograph (BOES) at Bohyunsan Optical Astronomy Observatory (BOAO) in Korea. The observations were made during 8 nights from 2018 January 20 to 2020 January 4 using 1.8 m telescope with a fiber having a diameter of 300 $\mu$m ($R$ $\sim$ 30000). The BOES covers a wavelength range of 3,600 – 10,200 {\AA} \citep{Kim+2007PASP..119.1052K}.  The exposure time was set at 1500 s. The mean signal-to-noise (S/N) ratio of the observed spectra was approximately 42 at 5000 {\AA}. Data reduction was performed using the CCDRED and ECHELLE packages in the IRAF.

\begin{figure}[]
\includegraphics[width=1\textwidth]{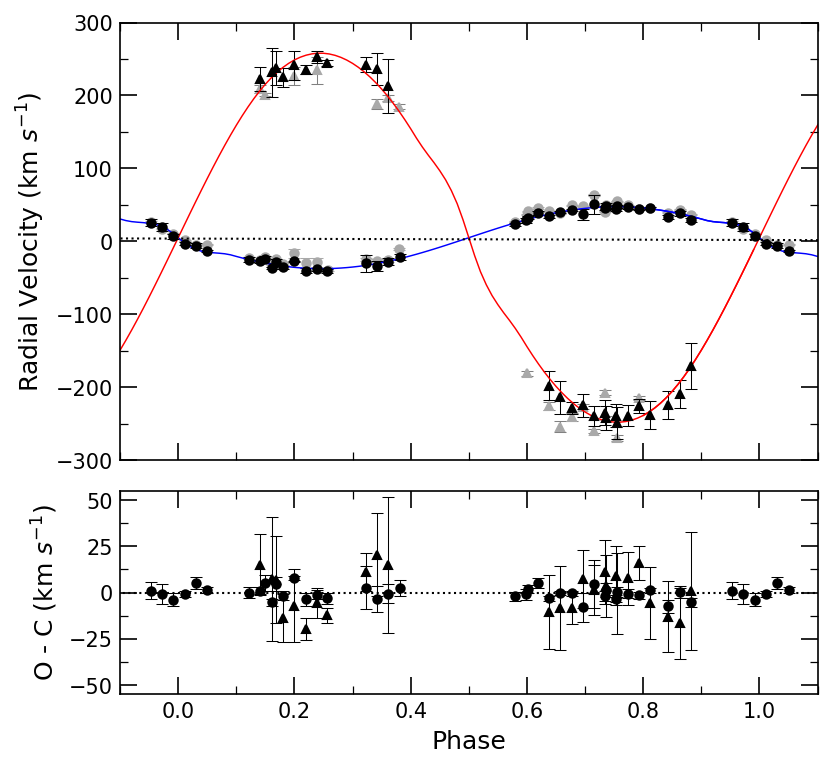}
\caption{The RVs of EW Boo with the synthetic curves. The circles and triangles represent the RV measurements for primary and secondary stars, respectively. The black and gray symbols represent the RVs of both components measured using two Gaussian functions and Broadening function, respectively. The solid curves were obtained from the light and RV solutions using the WD code. The dotted line denotes the system velocity of $+$5.09 km s$^{-1}$. In the lower panel, the residuals between observed RVs and theoretical models are plotted.
\label{fig:rv}}
\end{figure}

To measure the RVs, we searched for absorption lines that clearly separated both components in the entire wavelength range. It was observed that only the lines of Fe I $\lambda$4957 could identify the features of both the components. The RV measurements were performed in two ways: First, by using the broadening function method implemented in the RAVESPAN code\footnote{https://users.camk.edu.pl/pilecki/ravespan/} \citep{Pilecki+2017ApJ...842..110P} with the template spectra obtained from the synthetic spectral library of \citet{Coelho+2005A&A...443..735C}. And second, by fitting the absorption line for each binary component using two Gaussian functions of the IRAF splot task, following the procedure used by \citet{Hong+2015AJ....150..131H}. The measurements using two Gaussian functions were repeated several times, and the resulting RVs and their errors were adopted as the averages and standard deviations of the measured values, respectively. The measured RVs of both components and their errors are presented in Table~\ref{tab:RV} and plotted in Figure~\ref{fig:rv}. As shown in Figure~\ref{fig:rv}, the RVs measured using the Gaussian method conforms to the RVs measured using RAVESPAN. However, the RVs measured by RAVESPAN had errors that were negligible compared to the large scatter of the RV curves. In contrast, the Gaussian function method determined more RV points and estimated reasonable errors. Therefore, we used the RVs determined using Gaussian functions in our analysis.

To determine the atmospheric parameters of the binary components, we separated the blended spectra of the primary and secondary components using the FDBINARY spectral disentangling code \citep{Ilijic+2004ASPC..318..111I}, based on the Fourier transform of \citet{Hadrava_1995A&AS..114..393H}. Absorption lines were selected from four spectral regions:  Fe II $\lambda$4046, Fe I $\lambda$4271, Fe I $\lambda$4383, and Mg I $\lambda$4481, which are good temperature indicators of the A-F type main-sequence stars presented by the $Digital$ $Spectral$ $Classification$ $Atlas$ of R. O. Gray. In this process, the reference epoch ($T_{0}$), orbital period ($P$), and RV semi-amplitudes ($K_{1,2}$) are adopted from the binary parameters in Section~\ref{sec:4}. As a result, considering the light contribution of the secondary component was not sufficient to obtain a reliable model spectrum, we obtained only the disentangling spectrum of the primary component.

\begin{figure}[]
\plotone{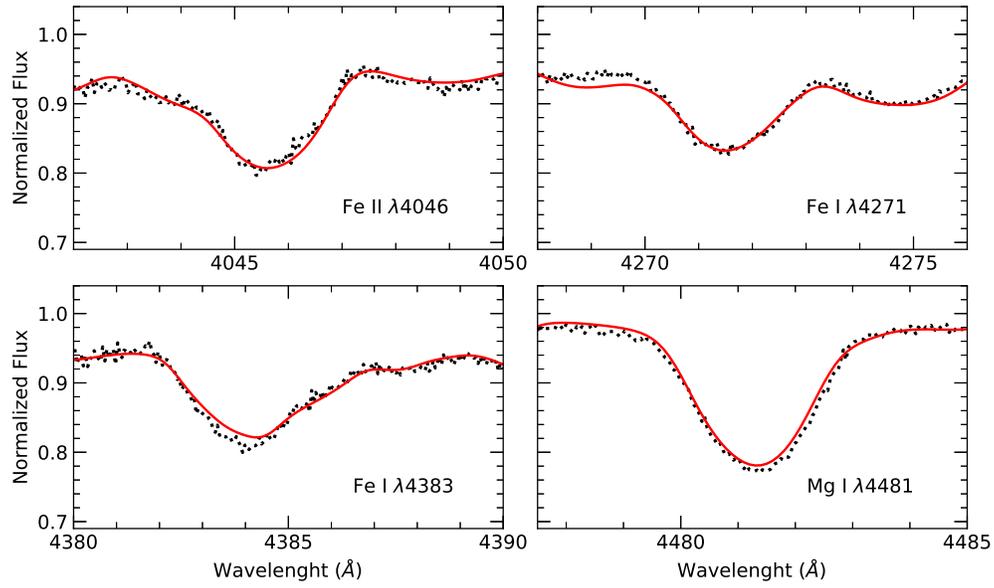}
\caption{The four spectral absorption lines of the primary component. The black dots and red solid lines represent the disentangling spectrum obtained by the FDBINARY code and the synthetic spectrum at 8560 K and 90 km s$^{-1}$, respectively. \label{fig:spectrum_fit}}
\end{figure}

We examined the atmospheric parameters by determining the theoretical spectra that best matched the disentangling spectrum of the primary component. For this purpose, theoretical spectra were generated using the PYTHON framework ISPEC \citep{Blanco-Cuaresma+2014A&A...566A..98B,Blanco-Cuaresma_2019MNRAS.486.2075B}, which included various synthetic spectrum calculation codes, model atmospheres, and line lists. During this process, we adopted SPECTRUM code \citep{Gray&Corbally_1994AJ....107..742G}, ATLAS-9 \citep{Castelli&Kurucz_2003IAUS..210P.A20C} model atmospheres and a third version of the Vienna Atomic Line Database \citep[VALD3;][]{Kupka+2011BaltA..20..503K}, respectively. Moreover, we assumed that the solar metallicity of [Fe/H] = 0.0. The micro- and macroturbulence velocities were calculated using the empirical relation based on GES UVES data release 1 \citep{Jofr+2014A&A...564A.133J,Blanco-Cuaresma+2014A&A...566A..98B}. The linear limb darkening coefficient was assumed to be $X_{1}$ = 0.561 and interpolated from the tables published by \citet{vanHamme_1993AJ....106.2096V}. The surface gravity was set to log $g_{1}$ = 4.23, as obtained in Section~\ref{sec:4}. A total of 15311 theoretical spectra were generated in the range of 6500 K $\leq T_{\rm{eff,1}} \leq$ 9000 K, and 50 km s$^{-1}$ $\leq v_1\sin i \leq$ 110 km s$^{-1}$ at intervals of 10 K and 1 km s$^{-1}$, respectively. Then, we simultaneously determined the global minimum for both the effective temperature ($T_{\rm{eff,1}}$) and the projected rotational velocity ($v_{1}$sin$i$) by calculating the $\chi^{2}$ statistics between the theoretical and four spectral regions of the disentangling spectrum. Lastly, the resulting best-fitting values of $T_{\rm{eff,1}} = 8560$ $\pm$ $118$ K and $v_{1}$sin$i$ = $90$ $\pm$ $9$ km s$^{-1}$ were determined. The disentangling spectrum of the primary component is shown in Figure~\ref{fig:spectrum_fit} along with the best-fit theoretical spectra. Our temperature corresponded to the spectral type of A3 \citep{Pecaut&Mamajek_2013ApJS..208....9P}, which is closer to A2 of DG than A7 of ZLW.

\section{Orbital Period Study} \label{sec:3}
From the TESS observations, a total of 113 times of minimum were determined using the method of \citet[][hereafter KW]{Kwee&vanWoerden_1956BAN....12..327K}. An additional 31 timings from the Super Wide Angle Search for Planets \citep[][hereafter SWASP]{Butters+2010A&A...520L..10B,Street+2003ASPC..294..405S} measurements were derived using the KW method. A total of 183 timings, including the TESS and SWASP timings, were collected from the literatures and database of \citet{Kreine+2001aocd.book.....K}. The HJD timings based on ground observations were converted into a TDB-based BJD using online applets \citep{Eastman+2010PASP..122..935E}\footnote{http://astroutils.astronomy.ohio-state.edu/time/}. Table~\ref{tab:timing} of Appendix~\ref{sec:appendixA} lists all the timings together with their errors.

\begin{figure}[]
\includegraphics[width=0.9\textwidth]{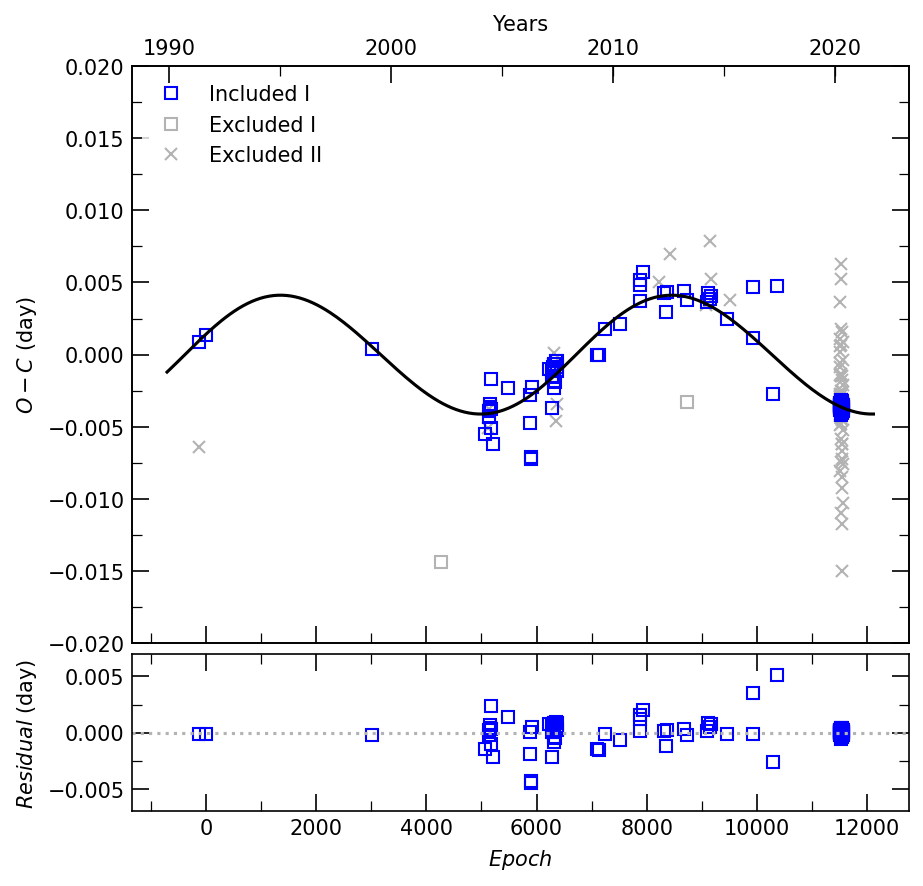}
\caption{The eclipse timing ($O-C$) diagram of EW Boo with the terms of LITE ephemeris. In the top panel, the solid curve represents the $\tau_3$ orbit. The bottom panel shows the residuals from the complete ephemeris.
\label{fig:3body_q}}
\end{figure}

The (O-C) residuals of all eclipse timings were calculated with \def\citeapos{Kreiner_2004AcA....54..207K}{\citeauthor{Kreiner_2004AcA....54..207K}'s (\citeyear{Kreiner_2004AcA....54..207K})} light elements and plotted in the upper panel of Figure~\ref{fig:3body_q}. As seen in Figure~\ref{fig:3body_q}, this eclipse timing diagram (ETD) shows that the orbital period of EW Boo has changed in a sinusoidal pattern rather than a secular increase suggested by ZLW. In addition, the most recent secondary timings obtained from the TESS observations showed severe deviations from the primary ones, considering that the shallow secondary eclipses were disturbed by pulsation. Therefore, we excluded all secondary timings from subsequent analyses. Additionally, we excluded two primary timings considering their residuals showed unreasonably large deviations compared to the neighboring ones. These timings are also excluded in subsequent analysis and marked with the superscript ``*'' in Table~\ref{tab:timing}. Assuming that the sinusoidal change shown in Figure 4 is due to the light-time effect (LITE) by a third body, 111 timings were fitted to the LITE ephemeris as follows:

$$C = T_{0}+PE+\tau_{3},\,\,\,\,\,\,\,\,\,\,\,\,\,\,\,\,\,\,\,\,\,\,\,\,\,\,\,\,(1)$$

\noindent where the LITE term ($\tau_{3}$) has five unknown parameters ($a_{12}$sin$i_{3}$, $e$, $\omega$, $n$, and $T$) whose details were given by \citet{Irwin_1952,Irwin_1959}. The Levenberg-Marquardt method \citep{Press+1992nrfa.book.....P} was applied to solve the unknown parameters, which included the initial epoch ($T_{0}$) and orbital period ($P$). The final solution parameters are presented in Table~\ref{tab:3bodyq}. In this calculation, each weight for all timings was inversely proportional to the squared value of the respective error for all timings. The errors of the six timings without error were set to an average value of 0.0009 d for all timings. The solid curve in the upper panel of Figure~\ref{fig:3body_q} represents $\tau_{3}$ in equation (1). The residuals from equation (1) were shown in the bottom panel of Figure~\ref{fig:3body_q}. As summarized in Table~\ref{tab:3bodyq}, the third body had an LITE period of 17.6 $\pm $ 0.3 years and a semi-amplitude of 0.0041 $\pm$ 0.0001 d. The mass of the outer companion was estimated to be 0.22 M$_{\odot}$, assuming that it has an orbit coplanar with the eclipsing pair of the EW Boo. Furthermore, assuming that the circumbinary object is a main-sequence star, it may contribute approximately 0.02$\%$ to the total light of the system.

\section{Binary modeling And Absolute dimensions} \label{sec:4}
Before the binary modeling, we checked whether there was another star around EW Boo to determine whether there were any factors that could distort the light curves. Gaia EDR3 1295253976312344704 (hereinafter VIS) is located 2 arcsec away from EW Boo \citep{Gaia+2020yCat.1350....0G}. Considering the pixel scale of TESS photometry is approximately 21 arcsec, EW Boo's TESS light curves naturally contain the light of VIS. The TESS magnitude of the nearby object was converted and found to be $T_{p}$ = 13.84 mag based on its Gaia magnitude \citep{Stassun+2019AJ....158..138S}, and its magnitude corresponded to approximately 3$\%$ of the total light contribution of the TESS data.

The TESS light and our double-lined RV curves were simultaneously analyzed using the 2015 version of the Wilson--Devinney (WD) binary model code \citep{Wilson&Devinney_1971ApJ...166..605W, Wilson&VanHamme_2014ApJ...780..151W}. The IPTMAX parameter (default: IPTMAX = 20,000) specifies the maximum number of data points that the WD 2015 code can handle. In order to process the TESS light curve consisting of a total of 36,550 points at once, the IPTMAX was set to 40,000 in the WD code, which was executed smoothly. Also, instead of phase, the BJD time of each measurement was used as the independent variable, allowing ephemeris parameters (initial epoch $T_0$ and orbital period $P$) to be determined by utilizing not just times of eclipse minima but whole light and/or radial velocity curves \citep{Wilson_2005Ap&SS.296..197W}. This idea was successfully applied to some EBs such as CN And \citep{VanHamme+2001AJ....122.3436V}, AR Mon \citep{Williamon+2005AJ....129.2798W}, and YY Cet \citep{Williamon+2012PASP..124..411W}.

The effective temperature of the hot primary star was set to $T_{\rm{eff,1}}$ = 8560 K, as obtained from our spectral analysis. The gravity darkening coefficients and bolometric albedo for the primary and secondary components were set to $g_{1}$ = 1.0, $g_{2}$ = 0.32, $A_{1}$ = 1.0, and $A_{2}$ = 0.5, respectively, based on their temperatures. The logarithmic monochromatic ($x$, $y$) and bolometric ($X$, $Y$) limb darkening coefficients were interpolated from \citet{vanHamme_1993AJ....106.2096V} who recently added the TESS passband (IBAND = 95) to his data files for limb darkening coefficients that need to be accessible when running the 2015 WD code.\footnote{https://faculty.fiu.edu/~vanhamme/lcdc2015/} The ratio of the axial rotation rate to the mean orbital rate was set to $F_{1, 2}$ = 1.0, for both components. During the analysis, although the differential correction (DC) program of the WD code began from mode 2 (detached system), it was always converged to mode 5 (semi-detached system with the secondary component filled with its inner Roche lobe) during computation. The adjusted parameters were the ephemeris epoch ($T_{0}$), orbital period ($P$), system velocity ($\gamma$), semi-major axis ($a$), orbital inclination ($i$), effective temperature of the secondary component ($T_{2}$), surface potential of the primary component ($\Omega_{1}$), mass ratio ($q$), and monochromatic luminosity ($L_{1}$). In addition, the third light ($l_{3}$) was adjusted to consider the light contribution of the neighboring star VIS around EW Boo. The results (Model 1) are listed in the second and third columns of Table~\ref{tab:binarymodel}.

\begin{figure}[]
\includegraphics[width=0.9\textwidth]{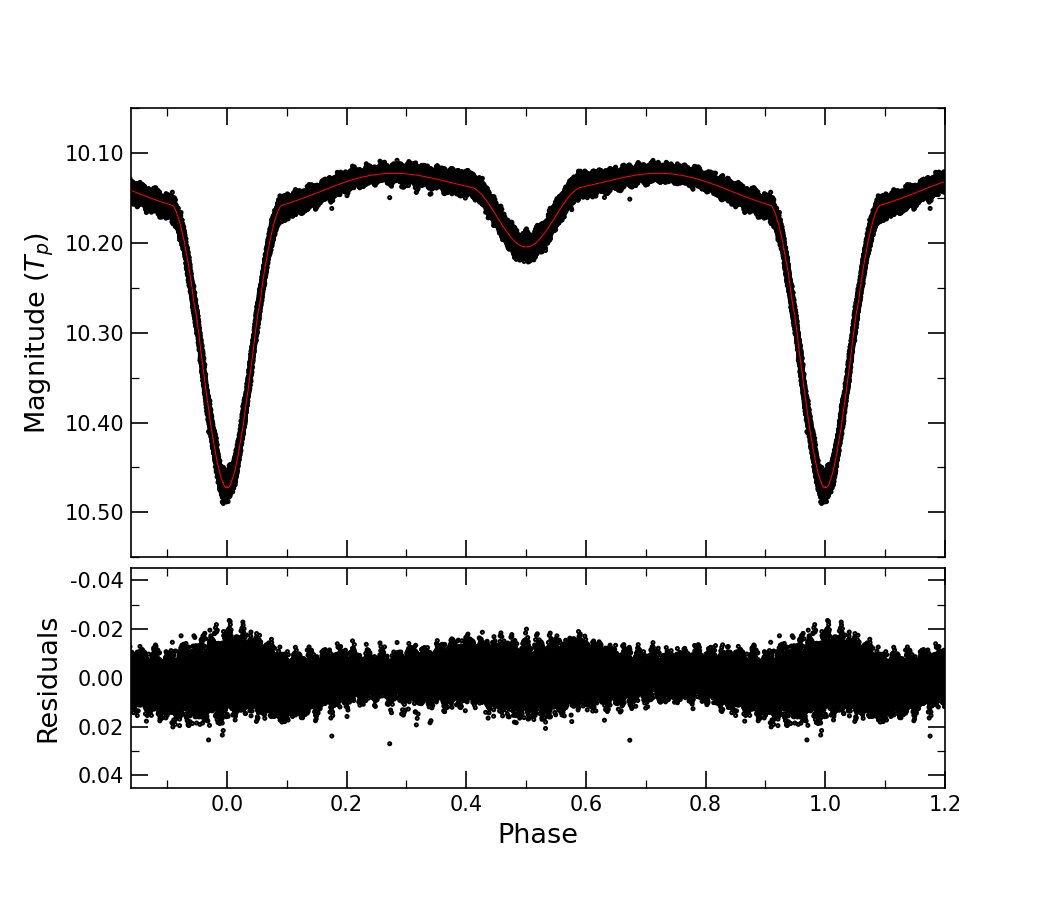}
\caption{The black circles represent the observed TESS data (top) and their corresponding residuals (bottom). The red synthetic curve in the top panel was obtained from Model 2 in Table~\ref{tab:binarymodel}. \label{fig:wd_lc}}
\end{figure}

Through spectral analysis using the orbital inclination, the rotational velocity of the primary star was determined as $v_{1}$ = $93$ $\pm$ 9 km s$^{-1}$. Considering the orbital period and its stellar radius, the synchronous rotational velocity of the primary component was calculated as $v_{\rm 1, sync}$ = $112$ $\pm$ 1 km s$^{-1}$. Thus, the rotational velocity of the primary star, estimated from the spectroscopic analysis, was 0.83 times slower than the synchronous rotational velocity. Therefore, the light and RV curves were simultaneously reanalyzed by setting the ratio of the axial rotation rate to the mean orbital rate to $F_{1}$ = $v_{1}/v_{\rm 1, sync}$ = 0.83. The results are listed in Model 2 in the fourth and fifth columns of Table~\ref{tab:binarymodel}. In Table~\ref{tab:binarymodel}, there are no remarkable differences between Models 1 and 2, except that the dimensionless potential of Model 1 is slightly larger than that of Model 2. Moreover, the values of $\Sigma W(O-C)^2$ in both models are consistent. Naturally, we selected Model 2 as the final solution of EW Boo, because it is strongly supported by spectral analysis. Such asynchronous rotation for Algol type stars exhibiting circular orbits has also been shown in AB Cas \citep{Hong+2017AJ....153..247H}, V392 Ori \citep{Hong+2019AJ....157...28H}, V404 Lyr \citep{Lee+2020AJ....159...24L}, and TZ Dra \citep{Kahraman_Alicavus+2022MNRAS.510.1413K}. The TESS light curve phased with the ephemeris in Model 2 is shown in the upper panel of Figure~\ref{fig:wd_lc}, where the red curve denotes the synthetic light curve calculated with Model 2. The residuals from Model 2 are also plotted in the bottom panel; they exhibit a large scatter due to pulsation. As suggested by ZLW and DG, our solution also confirms that EW Boo is a pulsating Algol-type binary system with a secondary component filling its inner Roche lobe. The pulsational characteristics are discussed in the next section.

Comparing our solution (Model 2) and the DG’s, the parameters determined from light curves are almost identical in both models, except that the parameters (such as semi-major axis and system velocity) measured from radial-velocity curves are notably different between the two solutions. In the last section, we investigate whether these differences arise because of LITE by a tertiary body, as proposed in Section~\ref{sec:3}. 

The JKTABSDIM code \citep{Southworth+2005A&A...429..645S} was used to determine the absolute dimensions of EW Boo. The binary parameters ($P$, $i$, $r_{1,2}$, $K_{1,2}$, $T_{1,2}$) and their errors were set to the values of Model 2, as shown in Table~\ref{tab:binarymodel}. The absolute dimensions and their errors in addition to those obtained independently by ZLW and DG are listed in Table~\ref{tab:Absolutedimension}. As shown in Table~\ref{tab:Absolutedimension}, our absolute dimensions were more similar to those of DG than ZLW. This indicates that the parameters of our method and DG's were obtained from both the light and RV curves, whereas the parameters of ZLW were determined only from the light curves. Furthermore, the parameters of this study and DG differed from each other at a higher level than their uncertainties. The ratios of the DG parameters to ours for the primary and secondary components in mass, radius, and luminosity were approximately (0.75, 0.76), (0.94, 0.91), and (1.04, 0.98), respectively.

\section{Pulsational characteristics} \label{sec:5}
According to ZLW and DG, the pulsation frequencies detected from ground-based observations can be attributed to $\delta$ Sct-type pulsations. In addition, considering the binary parameters in Section 4, the primary star of EW Boo was located inside the $\delta$ Sct instability strip. Therefore, we intensively investigated the pulsation characteristics of the primary star with more reliable light residuals obtained by eliminating the eclipse effects from the TESS light curves as follows: First, the TESS photometric measurements were segmented into 12 data sets at intervals that are five times the orbital period. Second, each dataset was individually analyzed with the WD code by adjusting the reference epoch ($T_{0}$) and monochromatic luminosity ($L_{1}$) only, among the binary parameters of Model 2 listed in Table~\ref{tab:binarymodel}. Finally, the light residuals between each dataset and the theoretical dataset constructed by Model 2 with the corresponding $T_{0}$ and $L_{1}$ values were collected and integrated into one light residual dataset to probe the pulsating properties of the primary star.

\begin{figure}[]
\plotone{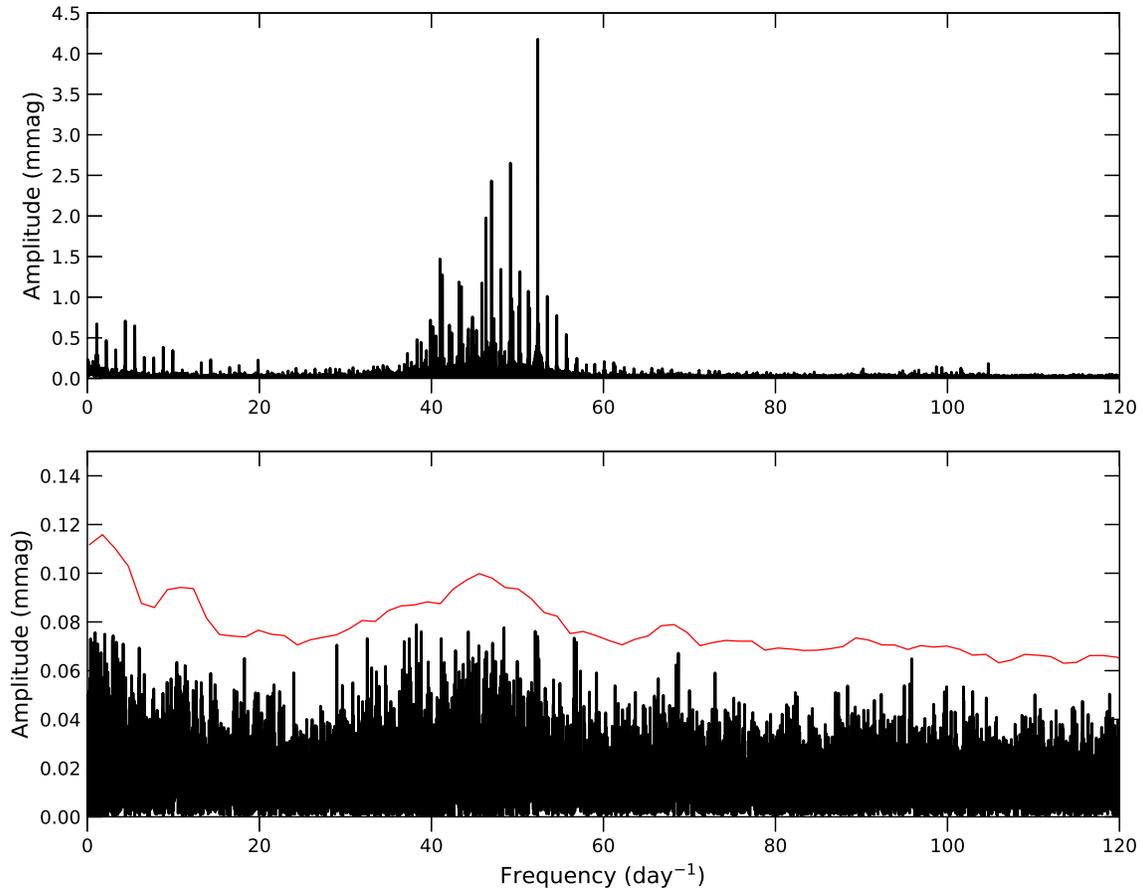}
\caption{Amplitude spectrum before (upper panel) and after (bottom panel) pre-whitening from the PERIOD04 program. The red solid line in the bottom panel is four times the local noise spectrum, which was calculated for the grids on 3 d$^{-1}$.
\label{fig:snr_freq}}
\end{figure}

\begin{figure}[]
\plotone{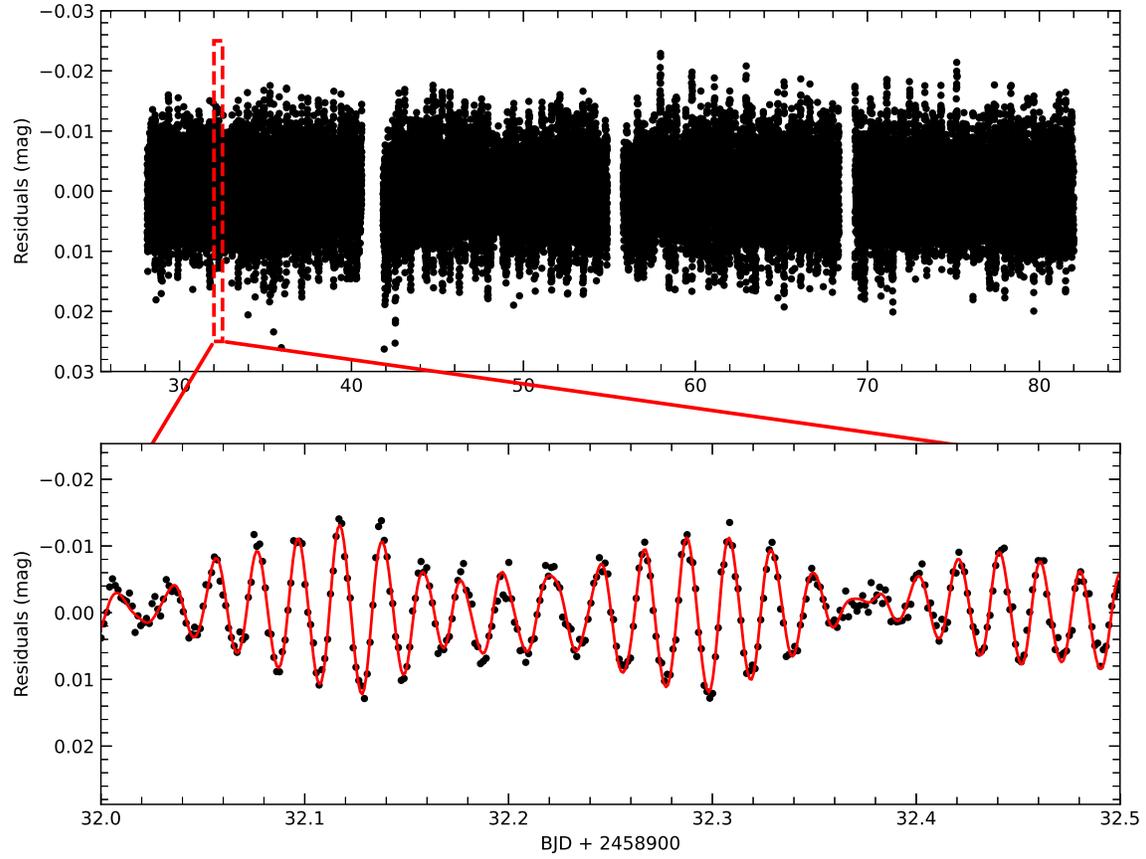}
\caption{The upper panel shows the time domain residuals of light curves subtracting binary effects. The lower panel shows a short section of the residuals in the inset box of the upper panel. The synthetic curve was computed from the 127 frequency fit to the data.
\label{fig:pul_draw}}
\end{figure}

Considering the light of the primary star was blocked by the secondary star during the primary eclipses, the out-of-primary eclipse part of the integrated light residuals was used in our subsequent analysis. Multiple frequency analysis was carried out in the frequency range from 0 to the Nyquist limit of 360 d$^{-1}$ using the Fourier analysis package PERIOD04 \citep{Lenz&Breger2005CoAst.146...53L} using the procedure described by \citet{Lee+2014AJ....148...37L}. A total of 127 frequencies were detected with a significance criterion of S/N $>$ 4.0 in amplitude \citep{Breger+1993A&A...271..482B}. The amplitude spectra of EW Boo before and after prewhitening all 127 frequencies are illustrated in the upper and bottom panels of Figure~\ref{fig:snr_freq}, respectively. The resultant frequencies, amplitudes, and phases of the pulsations are presented in Table~\ref{tab:freq_all} in Appendix~\ref{sec:appendixB}, where the errors were calculated using the equations in \citet{Kallinger+2008A&A...481..571K}. The synthetic curve calculated from the 127-frequency fit is represented by a red curve in the lower panel of Figure~\ref{fig:pul_draw}.

We searched for possible harmonic and combination frequencies within frequency resolution \citep{Loumos&Deeming_1978Ap&SS..56..285L} of $\Delta f$ = 1.5/$\Delta T$ = 0.028 d$^{-1}$, where the time span of the TESS observations is $\Delta T$ = 53.9 d. The frequencies of $f_{11}$, $f_{12}$, $f_{16}$, $f_{43}$, and $f_{54}$ were identified as multiple harmonics of the orbital frequency ($f_{\rm orb} = 1.10335$ d$^{-1}$). The candidates for the combination terms for the other frequencies are marked in the last column of Table~\ref{tab:freq_all}. A total of 19 frequencies did not correspond to the harmonic and combination frequencies, as shown in Table~\ref{tab:indi_freq} along with their pulsation constants, which were calculated using the equation log $Q_{i}$ = $-$ log $f_{i}$ + $0.5$ log $g$ + $0.1$ $M_{\rm bol}$ + log $T_{\rm eff}$ $-$ $6.456$ \citep{Petersen&Jorgensen_1972A&A....17..367P}. Without $f_{28}$ and $f_{30}$, their $Q$ values clustered between 0.0109–0.0141 d, which correspond to the pressure modes of the $\delta$ Sct pulsators with $Q$ $<$ 0.04 d \citep{Breger_2000ASPC..210....3B}.

As described in Section~\ref{sec:1}, ZLW and DG obtained four ($f_{1}$ = 52.372, $f_{2}$ = 47.320, $f_{3}$ = 49.481, and $f_{4}$ = 50.296) and two ($f_{1}$ = 49.4496 and $f_{2}$ = 43.8991) pulsation frequencies, respectively. Comparing these frequencies with ours in Table~\ref{tab:indi_freq}, ZLW’s $f_{1}$, $f_{2}$ and $f_{3}$ correspond to our $f_{1}$, $f_{10}$ and $f_{15}$, respectively. DG’s $f_{1}$ and $f_{2}$ are similar to our $f_{15}$ and $f_{9}$, respectively. Therefore, the frequencies in Table~\ref{tab:indi_freq} provide a more detailed picture of the pulsation state of EW Boo, confirming and including the frequencies found in previous studies.

To identify pulsation modes of EW Boo, each of the $Q$ values in Table 5 was compared with each of the theoretical $Q$ values of lots of submodes for radial ($l$ = 0) or non-radial pulsations ($0$ $<$ $l$ $\leq$ $3$) with 2.5 M$_{\odot}$ model tabulated by \citet{Fitch_1981ApJ...249..218F}, and the difference ($\triangle Q$) between the ones that matched well was found. The pulsation mode and difference in the $Q$ values for each frequency are listed in the seventh to tenth columns of Table~\ref{tab:indi_freq}, where $N$ denotes the $N$th order overtone mode for radial pulsation or the $N$th degree $p$-mode for non-radial pulsation. It shows that $\triangle Q$s of non-radial pulsation modes are approximately 1.7 to 26 times smaller than the corresponding values of radial pulsation modes, indicating most of the EW Boo pulsations may result from non-radial pulsations in 4 to 6 degree $p$-modes.

To crosscheck the frequencies, an additional frequency analysis was performed using the pre-search data conditioning SAP (PDCSAP) flux data, provided by the Science Processing Operations Center (SPOC) pipeline in target pixel files (TPFs), with long-term trends eliminated from the SAP flux using co-trending basis vectors (CBVs). The details of the SPOC pipeline were provided by \citet{Jenkins+2016SPIE.9913E..3EJ}. The binary effects of the PDCSAP data were eliminated using our binary model (Model 2). Following this, we compared the frequencies in Table~\ref{tab:indi_freq} to those detected from the PDCSAP data. 19 frequencies were detected in both datasets (SAP and PDCSAP), 17 of which corresponded to $\delta$ Sct pulsations, whereas the remaining two frequencies, possibly caused by the removal of the binary effect, were comparable to the orbital frequency ($f_{\rm orb} = 1.10335$ d$^{-1}$).

\section{summary and discussion} \label{sec:6}
The physical properties of EW Boo were comprehensively studied using our high-resolution spectroscopy and TESS photometry. From the time-series spectral analysis, we obtained new double-lined RV curves and determined the atmospheric parameters of $T_{\rm{eff,1}} = 8560$ $\pm$ $118$ K and $v_{1}$sin$i$ = $90$ $\pm$ $9$ km s$^{-1}$. The orbital period variation was first found as a periodic modulation with a cycle of 17.6 $\pm $ 0.3 years and a semi-amplitude of 0.0041 $\pm$ 0.0001 d. Owing to a tertiary companion with a nearly circular orbit of $e$ = 0.04 $\pm$ 0.04, the LITE was considered to be a cause of the sinusoidal variation. If the orbit is coplanar with the eclipsing pair ($i$ = 74$^{\circ}$.34), the mass of the suggested third body is calculated as 0.22 M$_{\odot}$. Following the empirical relations of \citet{Southworth_2009MNRAS.394..272S}, it was estimated to be of M5 type \citep{Pecaut&Mamajek_2013ApJS..208....9P} star with a $T_{3}$ = 3017 K and $R_{3}$ = 0.23 R$_{\odot}$.

The simultaneous analysis of the light and RV curves showed that EW Boo is a semi-detached system with $i$ = 74$^{\circ}$.34, $q$ = 0.160, and a temperature difference of $(T_{1}-T_{2})$ = 3732 K. In our binary modeling, the third light ($l_{3}$) was yielded at approximately 0.7$\%$, whose main source was VIS, which is only two arcsec away from the EW Boo. The absolute dimensions of the primary and secondary components were calculated as $M_{1}$ = 2.67 $\pm $ 0.08 M$_{\odot}$, $M_{2}$ = 0.43 $\pm $ 0.01 M$_{\odot}$, $R_{1}$ = 2.01 $\pm $ 0.02 R$_{\odot}$, $R_{2}$ = 1.35 $\pm $ 0.01 R$_{\odot}$, $L_{1}$ = 19.6 $\pm $ 0.5 L$_{\odot}$, and $L_{2}$ = 0.90 $\pm $ 0.05 L$_{\odot}$, respectively. Multi-frequency analysis was applied to the residuals of the TESS photometric light curves with the binary effects removed. A total of 127 frequency signals with S/N $>$ 4.0 were detected in the light residuals using out-of-primary-eclipse data. Among these, we acquired 19 independent frequencies, except for the possible combination frequencies and orbital multiples within the frequency resolution. These frequencies were also confirmed by the PDCSAP data analysis.

\begin{figure}[]
\includegraphics[width=0.9\textwidth]{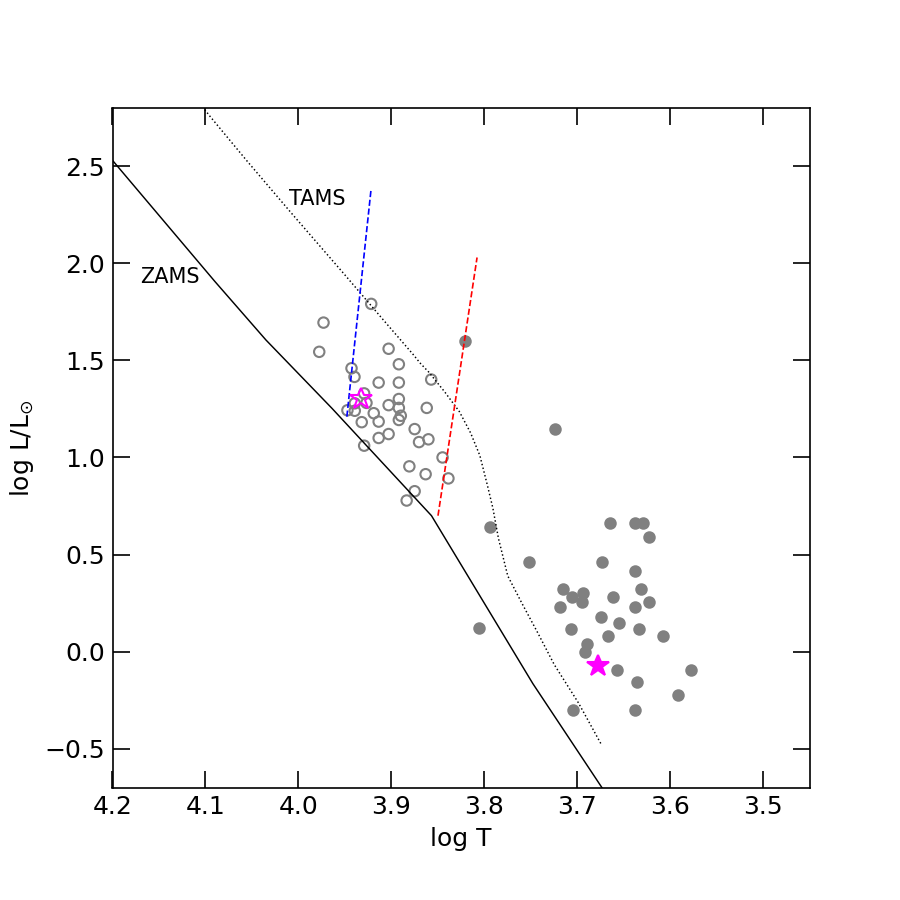}
\caption{The HR diagram for EW Boo (star symbols) and other semi-detached EBs with a $\delta$ Sct component \citep[circles;][]{Kahraman_Alicavus+2017MNRAS.470..915K}. The primary and secondary components for EW Boo are marked as the open and filled symbols, respectively. The blue and red lines represent the theoretical edges of the $\delta$ Sct instability strip \citep{Rolland+2002CoAst.142...57R, Soydugan+2006MNRAS.366.1289S}. The black solid and dotted lines denote the ZAMS and TAMS lines \citep{Girardi+2000A&AS..141..371G}.
\label{fig:hrd}}
\end{figure}

The Hertzsprung-Russell (HR) diagram in Figure~\ref{fig:hrd} shows the evolutionary status of both EW Boo components in addition to the 38 semi-detached EBs with a $\delta$ Sct component, as listed in \citet{Kahraman_Alicavus+2017MNRAS.470..915K}. Here, the more massive primary star is located between the ZAMS and TAMS lines, near the blue edge of the instability strip. In contrast, the less massive secondary star is located above the TAMS line along with the other secondaries. As a result, EW Boo comprises the oscillating primary star, with a spectral type of A3V, that fills approximately 70 $\%$ of the inner Roche lobe and the secondary star that fills the entire inner Roche lobes. This indicates that the initially more massive stars transferred most of their mass to the gainer, evolving into less massive secondary stars and the gainer developed into the present primary star in the $\delta$ Sct instability region \citep{Sarna_1993MNRAS.262..534S, Erdem&Ozturk_2014MNRAS.441.1166E}.

It was noted in Section~\ref{sec:4} that a significant difference in semi-major axis ($a$) and system velocity ($\gamma$) between our spectroscopic solutions and that of DG: ($a$, $\gamma$)=(5.739 $\pm$ 0.003 R$_\odot$, 5.04 $\pm$ 0.05 km s$^{-1}$) for our solution, and (5.221 $\pm$ 0.043 R$_\odot$, 17.923 $\pm$ 0.152 km s$^{-1}$) for the DG one. The difference in system velocity at the two epochs, with an interval of about 7 years, is approximately 12.9 km s$^{-1}$, which is significantly larger than the calculated standard error. We investigated whether the difference could be due to the orbital motion of the eclipsing pair's center of mass around the mass center of the triple system. By using the LITE parameters in Table~\ref{tab:3bodyq}, the semi-amplitude of the radial velocity of the center of mass of the eclipsing pair was calculated to be, at most, as 1.2 km s$^{-1}$ which is too small to be the cause of the difference. Consequently, the significant difference in system velocity presently remains a daunting issue.

The Gaia parallaxes of EW Boo and VIS are $2.30 \pm 0.02$ mas and $2.24 \pm 0.06$ mas, respectively, which are consistent within the error. Additionally, the proper motion for EW Boo ($\mu_{\rm RA} = 31.520 \pm 0.01$ mas yr$^{-1}$, $\mu_{\rm Dec} = -18.31 \pm 0.02$ mas yr$^{-1}$) were similar to those of the VIS ($\mu_{\rm RA} = 31.80 \pm 0.05$ mas yr$^{-1}$, $\mu_{\rm Dec} = -18.57 \pm 0.07$ mas yr$^{-1}$). This may imply three possibilities: (1) VIS is the cause for the variation in the ETD of EW Boo; (2) While VIS is one component of a wide multiple system that includes the EW Boo pair, it does not cause variations in the ETD; (3) VIS and EW Boo coincidentally have similar proper motions but are independent stars. To investigate whether VIS can be the cause of the variation in the ETD, we calculated the distance between VIS and EW Boo. The angular distance between the EW Boo and VIS was calculated to be 2 arcsec based on RA and Dec provided by Gaia EDR3. Using the Gaia parallax, the Gaia distance to EW Boo and the VIS were determined to be $434 \pm 3$ pc and $447 \pm 12$ pc, respectively. Accordingly, the distance between VIS and EW Boo was calculated to be approximately 13 pc ($= 4.01 \times 10^{14}$ km) by using the law of cosine based on the Gaia distances of EW Boo and the VIS, and the angular distance between them. Compared to $a_{12} \sin i_{3} $ of a third body (see Table~\ref{tab:3bodyq}), VIS is located too far away to cause variations in the ETD. However, because the typical orbital periods of wide binaries are significantly long such as decades or longer, continuous astrometric observations are required to confirm whether EW Boo and VIS are gravitationally bound.

\acknowledgments
We thank the anonymous reviewer for his (or her) careful comments and suggestions, which have enhanced the original paper.
This paper includes data collected by the TESS mission, which were obtained from MAST. Funding for the TESS mission is provided by the NASA Explorer Program. The authors wish to thank the TESS team for its support of this work. This research has made use of the SIMBAD database, operated at CDS, Strasbourg, France. C.-H. K. was supported by grants of the National Research Foundation of Korea (2019R1A2C2085965 and 2020R1A4A2002885). J.W.L. was supported by the KASI grant 2022-1-830-04.

\bibliography{ewboo_draft}{}
\bibliographystyle{aasjournal}

\begin{deluxetable}{rrrrrrrrrr}
\tablewidth{0pt}
\tablecaption{Radial Velocities of EW Boo \label{tab:RV}}
\tablehead{
\colhead{} & \colhead{}& \multicolumn{4}{c}{Gaussian Fitting} & \multicolumn{4}{c}{RAVESPAN}\\
\cline{3-10}
\colhead{BJD} & \colhead{Phase} & \colhead{$V_{1}$}& \colhead{$\sigma_{1}$} & \colhead{$V_{2}$}& \colhead{$\sigma_{2}$}& \colhead{$V_{1}$}& \colhead{$\sigma_{1}$} & \colhead{$V_{2}$}& \colhead{$\sigma_{2}$}\\
\colhead{(2458000+)} & \colhead{} & \colhead{(\rm{km s}$^{-1}$)}& \colhead{(\rm{km s}$^{-1}$)} & \colhead{(\rm{km s}$^{-1}$)}& \colhead{(\rm{km s}$^{-1}$)}& \colhead{(\rm{km s}$^{-1}$)}& \colhead{(\rm{km s}$^{-1}$)} & \colhead{(\rm{km s}$^{-1}$)}& \colhead{(\rm{km s}$^{-1}$)}
}
\startdata
139.3088 & 0.734 & 45.4 & 2.8 & $-$234.6 & 17.4 & 40.8 & 0.9 & $-$207.2 & 2.9 \\
139.3265 & 0.754 & 44.2 & 2.0 & $-$238.8 & 15.9 & 51.9 & 0.9 & - & - \\
139.3441 & 0.773 & 46.6 & 2.5 & $-$238.8 & 14.5 & 49.5 & 1.0 & - & - \\
139.3617 & 0.793 & 44.7 & 1.9 & $-$225.6 & 9.1 & 43.8 & 1.0 & $-$214.8 & 2.6 \\
139.3794 & 0.812 & 45.9 & 1.4 & $-$238.3 & 19.4 & 46.0 & 1.0 & - & - \\
163.2247 & 0.122 & $-$25.0 & 3.2 & - & - & $-$22.5 & 1.0 & - & - \\
163.2424 & 0.141 & $-$27.5 & 2.0 & 222.3 & 16.3 & $-$25.5 & 1.0 & 209.0 & 5.1 \\
163.2600 & 0.161 & $-$36.6 & 1.9 & 231.9 & 33.6 & $-$33.5 & 1.1 & - & - \\
163.2776 & 0.180 & $-$35.4 & 2.2 & 224.6 & 13.0 & $-$30.9 & 1.9 & - & - \\
163.2953 & 0.200 & $-$27.6 & 1.0 & 241.5 & 19.7 & $-$16.5 & 5.8 & 226.7 & 12.7 \\
163.3129 & 0.219 & $-$40.3 & 3.5 & 235.5 & 6.0 & $-$29.1 & 7.0 & - & - \\
163.3306 & 0.239 & $-$38.5 & 2.4 & 252.4 & 8.2 & $-$28.5 & 6.2 & 234.5 & 18.7 \\
163.3472 & 0.257 & $-$40.4 & 2.9 & 245.1 & 4.1 & $-$38.7 & 1.9 & - & - \\
227.9896 & 0.580 & 23.3 & 2.5 & - & - & 26.8 & 0.9 & - & - \\
228.0072 & 0.599 & 28.7 & 3.0 & - & - & 36.4 & 0.9 & $-$180.7 & 3.7 \\
228.0248 & 0.619 & 39.0 & 2.6 & - & - & 45.4 & 1.1 & - & - \\
228.0425 & 0.638 & 34.1 & 1.5 & $-$198.1 & 19.9 & 41.2 & 0.9 & $-$225.8 & 5.8 \\
228.0601 & 0.658 & 40.2 & 1.1 & $-$213.8 & 22.7 & 38.8 & 0.8 & $-$253.9 & 7.7 \\
228.0777 & 0.677 & 42.5 & 1.3 & $-$228.4 & 8.7 & 49.2 & 0.9 & $-$240.7 & 5.5 \\
228.0954 & 0.697 & 37.1 & 7.9 & $-$224.8 & 15.6 & 48.0 & 0.8 & $-$226.9 & 3.5 \\
228.1130 & 0.716 & 51.0 & 13.0 & $-$239.3 & 13.6 & 63.4 & 1.1 & $-$260.2 & 2.6 \\
228.1306 & 0.735 & 48.5 & 3.7 & $-$242.4 & 16.7 & 50.1 & 0.9 & $-$242.5 & 2.5 \\
228.1482 & 0.755 & 47.9 & 2.6 & $-$248.5 & 21.9 & 55.1 & 0.9 & $-$269.1 & 4.1 \\
229.2339 & 0.953 & 25.6 & 4.6 & - & - & 26.6 & 0.8 & - & - \\
229.2516 & 0.972 & 19.5 & 5.6 & - & - & 17.6 & 1.0 & - & - \\
229.2692 & 0.992 & 6.8 & 3.4 & - & - & 10.6 & 0.9 & - & - \\
229.2868 & 0.011 & $-$2.9 & 1.7 & - & - & 1.3 & 0.9 & - & - \\
229.3044 & 0.031 & $-$6.0 & 3.2 & - & - & $-$5.2 & 0.8 & - & - \\
229.3222 & 0.050 & $-$13.3 & 1.6 & - & - & $-$4.4 & 0.9 & - & - \\
252.9769 & 0.149 & $-$24.7 & 4.4 & - & - & $-$21.7 & 0.9 & 201.0 & 2.4 \\
252.9945 & 0.169 & $-$27.7 & 3.6 & 237.7 & 23.5 & $-$24.7 & 1.2 & - & - \\
582.1329 & 0.322 & $-$30.2 & 11.6 & 241.9 & 10.4 & $-$24.7 & 1.8 & - & - \\
582.1505 & 0.342 & $-$33.8 & 6.9 & 236.5 & 22.3 & $-$26.6 & 2.2 & 187.8 & 6.7 \\
582.1682 & 0.361 & $-$27.8 & 5.1 & 212.8 & 36.8 & $-$26.0 & 1.9 & 196.3 & 3.7 \\
582.1858 & 0.381 & $-$21.2 & 4.3 & - & - & $-$10.4 & 1.0 & 184.5 & 3.5 \\
636.9851 & 0.843 & 33.5 & 3.3 & $-$224.1 & 19.3 & 39.5 & 1.0 & - & - \\
637.0027 & 0.863 & 38.3 & 3.1 & $-$209.7 & 19.0 & 43.3 & 0.9 & - & - \\
637.0205 & 0.882 & 29.2 & 2.4 & $-$170.4 & 31.7 & 36.5 & 0.9 & - & - \\
853.3803 & 0.602 & 32.3 & 2.1 & - & - & 41.1 & 0.9 & - & - \\
\enddata
\end{deluxetable}

\newpage

\begin{deluxetable}{lc}
\tablewidth{0pt}
\tablecaption{Parameters for the LITE ephemeris of EW Boo \label{tab:3bodyq}}
\tablehead{
\colhead{Parameter} & {Value}}
\startdata
$T_{0}$ (BJD) & 2448500.5145(5) \\
$P$ (d)  & 0.90634930(4) \\
$a_{12} \sin i_{3} $ (km)  & $1.07(2) \times 10^{8}$  \\
$\omega$ (deg)  & 358(115)        \\
$e$      & 0.04(4)       \\
$T$ (BJD) & 2448123(1851) \\
$P_{3}$ (yr)  & 17.6(3)        \\
$K$ (d)   & 0.0041(1)      \\
$\sigma_{\rm all}$ & 0.001 \\
\enddata
\tablecomments{$\sigma_{\rm all}$ is a rms scatter of all timing residuals}
\end{deluxetable}

\begin{deluxetable}{lcccc}
\tablewidth{0pt}
\tablecaption{Binary Parameters of EW Boo \label{tab:binarymodel}}
\tablehead{
\colhead{} & \multicolumn{2}{c}{Model 1 (F$_{1}$ = 1.00)} & \multicolumn{2}{c}{Model 2 (F$_{1}$ = 0.83)}  \\
\cline{2-3}
\cline{4-5}
\colhead{Parameter} & \colhead{Primary} & \colhead{Secondary} & \colhead{Primary} & \colhead{Secondary}
}
\startdata
$T_0$ (BJD) & \multicolumn{2}{c}{2458928.96936(3)} & \multicolumn{2}{c}{2458928.96655(3)} \\
$P$ (d) & \multicolumn{2}{c}{0.9063339(2)} & \multicolumn{2}{c}{0.9063339(2)} \\
$a$ (R$_{\odot}$)      & \multicolumn{2}{c}{5.736(3)} & \multicolumn{2}{c}{5.739(3)} \\
$K_1$ (km s$^{-1}$)         & \multicolumn{2}{c}{42.6(8)}     & \multicolumn{2}{c}{42.7(8)} \\
$K_2$ (km s$^{-1}$)         & \multicolumn{2}{c}{266(3)}  & \multicolumn{2}{c}{266(3)}   \\
$\gamma$ (km s$^{-1}$)      & \multicolumn{2}{c}{5.05(5)}    & \multicolumn{2}{c}{5.04(5)}\\
$q$ & \multicolumn{2}{c}{0.1602(3)} & \multicolumn{2}{c}{0.1605(3)}\\
$i$ (deg) & \multicolumn{2}{c}{74.38(1)} & \multicolumn{2}{c}{74.34(1)}\\
$T$ (K)     & 8560(118)       & 4820(150)  & 8560(118)      & 4828(150)   \\
$\Omega$          & 3.067(3)       & 2.130   & 3.044(3)       & 2.131     \\
$X, Y$     & 0.640, 0.239     & 0.643, 0.224  & 0.640, 0.239     & 0.643, 0.224   \\
$x, y$     & 0.712, 0.276     & 0.738, 0.260  & 0.712, 0.276     & 0.738, 0.260   \\
$l$/($l_{1}$+$l_{2}$+$l_{3}$)  & 0.9278(5)  & 0.0657   & 0.9273(5)      & 0.0659     \\
$l_{3}$ & \multicolumn{2}{c}{0.0065(5)} & \multicolumn{2}{c}{0.0068(5)}\\
$r$ (pole)  & 0.3428(3) & 0.2184(1) & 0.3456(3) & 0.2185(1)  \\
$r$ (point) & 0.3568(4) & 0.3218(1) & 0.3567(4) & 0.3220(1) \\
$r$ (side)  & 0.3514(3) & 0.2271(1) & 0.3514(3) & 0.2272(1)  \\
$r$ (back)  & 0.3546(3) & 0.2591(1) & 0.3545(3) & 0.2592(1)  \\
$\Sigma W(O-C)^2$ & \multicolumn{2}{c}{0.0025} & \multicolumn{2}{c}{0.0025} 
\enddata
\end{deluxetable}

\newpage

\begin{deluxetable}{lcccccc}
\tablewidth{0pt}
\tablecaption{Absolute Dimensions and Radiative Parameters of EW Boo \label{tab:Absolutedimension}}
\tablehead{
\colhead{} & \multicolumn{2}{c}{\citet{Zhang+2015AJ....149...96Z}} & \multicolumn{2}{c}{\citet{Dogruel+2015NewA...40...20D}} & \multicolumn{2}{c}{This paper}  \\
\cline{2-3}  \cline{4-5} \cline{6-7}
\colhead{Parameter} & \colhead{Primary} & \colhead{Secondary} & \colhead{Primary} & \colhead{Secondary} & \colhead{Primary} & \colhead{Secondary}}
\startdata
$M$ ($M_{\odot}$) & 1.80(18)  & 0.23(2) & 2.00(4)  & 0.327(8) & 2.67(8)  & 0.43(1)    \\
$R$ ($R_{\odot}$) & 1.84(6)  & 1.11(4) & 1.88(2)  & 1.23(1) & 2.01(2)  & 1.35(1)    \\
$L$ ($L_{\odot}$) & 10.9(5)  & 0.4(1) & 20.46(36)  & 0.88(2) & 19.63(47)  & 0.90(5)   \\
$T_{\rm eff}$ (K) & 7840  & 4514(35) & 8970  & 5042(24) & 8560(118)  & 4828(150)    \\
$\log$ $g$ (cgs) & 4.1(1) & 3.7(1) & 4.19(1) & 3.77(1) & 4.256(4) & 3.808(8)   \\
\enddata
\end{deluxetable}

\begin{deluxetable}{lccccccccc}
\tablewidth{0pt}
\tablecaption{The independent multiple frequencies of EW Boo \label{tab:indi_freq}}
\tablehead{\multirow{2}{*}{} & \multirow{2}{*}{Frequency (d$^{-1}$)} &  \multirow{2}{*}{Amplitude (mmag)} &  \multirow{2}{*}{Phase (rad)} &  \multirow{2}{*}{S/N} & \multirow{2}{*}{Q (d)} & \multicolumn{2}{c}{radial mode} & \multicolumn{2}{c}{non--radial mode}\\ \cline{7-8} \cline{9-10} & \colhead{} &  \colhead{} &  \colhead{} &  \colhead{} & \colhead{} & \colhead{($l$, $N$)} & \colhead{$\Delta$Q $\times$ 100 (d)} & \colhead{($l$, $N$)} & \colhead{$\Delta$Q $\times$ 100 (d)} }
\startdata
$f_{1}$ & 52.37033 $\pm$ 0.00005 & 4.20 $ \pm $ 0.04 & 1.51 $ \pm $ 0.03 & 193.03 & 0.0109 & (0, 6) & 0.021 & (3, 6) & 0.004 \\
$f_{2}$ & 49.19283 $\pm$ 0.00007 & 3.31 $ \pm $ 0.04 & 1.70 $ \pm $ 0.04 & 141.36 & 0.0116 & (0, 6) & 0.049 & (1, 6) & 0.014 \\
$f_{4}$ & 46.33846 $\pm$ 0.00012 & 2.01 $ \pm $ 0.04 & 4.06 $ \pm $ 0.06 & 80.31 & 0.0123 & (0, 5) & 0.023 & (3, 5) & 0.002 \\
$f_{5}$ & 40.99743 $\pm$ 0.00011 & 1.88 $ \pm $ 0.04 & 2.63 $ \pm $ 0.06 & 87.41 & 0.0139 & (0, 4) & 0.046 & (3, 4) & 0.021 \\
$f_{7}$ & 41.27599 $ \pm $ 0.00018 & 1.13 $ \pm $ 0.04 & 5.70 $ \pm $ 0.09 & 51.69 & 0.0138 & (0, 4) & 0.056 & (3, 4) & 0.031 \\
$f_{8}$ & 49.30983 $\pm$ 0.00023 & 0.96 $ \pm $ 0.04 & 5.01 $ \pm $ 0.12 & 41.23 & 0.0116 & (0, 6) & 0.049 & (1, 6) & 0.014 \\
$f_{9}$ & 43.48223 $ \pm $ 0.00024 & 0.95 $ \pm $ 0.04 & 0.11 $ \pm $ 0.13 & 39.23 & 0.0131 & (0, 5) & 0.057 & (1, 5) & 0.026 \\
$f_{10}$ & 47.28187 $ \pm $ 0.00035 & 0.65 $ \pm $ 0.04 & 1.95 $ \pm $ 0.18 & 27.08 & 0.0121 & (0, 5) & 0.043 & (3, 5) & 0.018 \\
$f_{13}$ & 44.27615 $ \pm $ 0.00041 & 0.56 $ \pm $ 0.04 & 0.37 $ \pm $ 0.22 & 23.04 & 0.0129 & (0, 5) & 0.037 & (2, 5) & 0.017 \\
$f_{14}$ & 40.15337 $ \pm $ 0.00034 & 0.59 $ \pm $ 0.04 & 2.76 $ \pm $ 0.18 & 27.80 & 0.0142 & (0, 4) & 0.016 & (3, 4) & 0.009 \\
$f_{15}$ & 49.48904 $ \pm $ 0.00035 & 0.63 $ \pm $ 0.04 & 1.17 $ \pm $ 0.18 & 27.31 & 0.0115 & (0, 6) & 0.039 & (1, 6) & 0.024 \\
$f_{17}$ & 40.52944 $\pm$ 0.00031 & 0.67 $ \pm $ 0.04 & 5.59 $ \pm $ 0.16 & 30.81 & 0.0141 & (0, 4) & 0.026 & (3, 4) & 0.001 \\
$f_{18}$ & 38.32320 $ \pm $ 0.00032 & 0.64 $ \pm $ 0.04 & 1.68 $ \pm $ 0.17 & 29.40 & 0.0149 & (0, 4) & 0.054 & (2, 4) & 0.023 \\
$f_{19}$ & 46.48238 $ \pm $ 0.00049 & 0.49 $ \pm $ 0.04 & 1.28 $ \pm $ 0.26 & 19.48 & 0.0122 & (0, 5) & 0.033 & (3, 5) & 0.008 \\
$f_{20}$ & 42.35961 $ \pm $ 0.00052 & 0.43 $ \pm $ 0.04 & 3.52 $ \pm $ 0.28 & 18.17 & 0.0135 & (0, 4) & 0.086 & (1, 5) & 0.014 \\
$f_{24}$ & 40.33073 $\pm$ 0.00059 & 0.35 $ \pm $ 0.04 & 5.16 $ \pm $ 0.31 & 16.07 & 0.0141 & (0, 4) & 0.026 & (3, 4) & 0.001 \\
$f_{26}$ & 44.16750 $ \pm $ 0.00060 & 0.38 $ \pm $ 0.04 & 5.88 $ \pm $ 0.32 & 15.80 & 0.0129 & (0, 5) & 0.037 & (2, 5) & 0.017 \\
$f_{28}$ & 1.15419 $ \pm $ 0.00068 & 0.41 $ \pm $ 0.05 & 3.70 $ \pm $ 0.36 & 14.06 & 0.4940 & & & & \\
$f_{30}$ & 1.21733 $ \pm $ 0.00066 & 0.41 $ \pm $ 0.05 & 1.48 $ \pm $ 0.35 & 14.33 & 0.4683 & & & & \\
\enddata
\end{deluxetable}

\newpage

\clearpage

\begin{appendix} 
\setcounter{table}{0} \renewcommand{\thetable}{A.\arabic{table}}
\section{List of Timings of minimum light} \label{sec:appendixA}
\renewcommand\arraystretch{0.95}
\begin{longtable}{lccccccc}
\tablewidth{0pt}
\caption{All times of minimum light for EW Boo \label{tab:timing}}\\
\hline\hline
\colhead{BJD} & \colhead{Error} & \colhead{Epoch} & \colhead{$O-C$} & \colhead{Me$^{a}$} & \colhead{Type$^{b}$} & \colhead{Reference}\\
\hline
\endfirsthead
\caption{continued}\\
\hline\hline
\colhead{BJD} & \colhead{Error} & \colhead{Epoch} & \colhead{$O-C$} & \colhead{Me$^{a}$} & \colhead{Type$^{b}$} & \colhead{Reference}\\
\hline
\endhead
\hline
\endfoot
2448385.8549$^{*}$ & 0.0009 & $-$126.5 & $-$0.0064 & CCD & II & \citet{Kreine+2001aocd.book.....K}\\
2448386.3153 & 0.001 & $-$126.0 & 0.0008 & CCD & I & \citet{Kreine+2001aocd.book.....K}\\
2448500.5158 & - & 0.0 & 0.0013 & CCD & I & \citet{ESA_1997yCat.1239....0E}\\
2451225.9072 & 0.0014 & 3007.0 & 0.0004 & CCD & I & \citet{Diethelm_2001IBVS.5027....1D}\\
2452362.4545$^{*}$ & 0.0015 & 4261.0 & $-$0.0143 & CCD & I & \citet{Agerer+2003IBVS.5484....1A}\\
2453097.5126 & 0.0012 & 5072.0 & $-$0.0055 & CCD & I & \citet{Hubscher2005aIBVS.5643....1H}\\
2453139.6583$^{*}$ & 0.0015 & 5118.5 & $-$0.0051 & CCD & II & This paper (SWASP)\\
2453154.6138 & 0.0003 & 5135.0 & $-$0.0043 & CCD & I & This paper (SWASP)\\
2453155.5206 & 0.0002 & 5136.0 & $-$0.0039 & CCD & I & This paper (SWASP)\\
2453160.5056$^{*}$ & 0.0008 & 5141.5 & $-$0.0038 & CCD & II & This paper (SWASP)\\
2453165.4909 & 0.0003 & 5147.0 & $-$0.0034 & CCD & I & This paper (SWASP)\\
2453175.4605 & 0.0008 & 5158.0 & $-$0.0037 & CCD & I & This paper (SWASP)\\
2453184.5239 & 0.0005 & 5168.0 & $-$0.0038 & CCD & I & This paper (SWASP)\\
2453185.4323 & 0.0006 & 5169.0 & $-$0.0017 & CCD & I & This paper (SWASP)\\
2453194.4924 & 0.0004 & 5179.0 & $-$0.0051 & CCD & I & This paper (SWASP)\\
2453224.4008 & 0.0005 & 5212.0 & $-$0.0062 & CCD & I & This paper (SWASP)\\
2453475.4635 & 0.0012 & 5489.0 & $-$0.0023 & CCD & I & \citet{Hubscher+2005IBVS.5657....1H}\\
2453831.6563 & 0.0013 & 5882.0 & $-$0.0048 & CCD & I & This paper (SWASP)\\
2453832.5646 & 0.0003 & 5883.0 & $-$0.0028 & CCD & I & This paper (SWASP)\\
2453851.5935 & 0.0005 & 5904.0 & $-$0.0072 & CCD & I & This paper (SWASP)\\
2453852.5000 & 0.0009 & 5905.0 & $-$0.0071 & CCD & I & This paper (SWASP)\\
2453862.4747 & 0.0048 & 5916.0 & $-$0.0022 & CCD & I & \citet{Hubscher+2006IBVS.5731....1H}\\
2454150.6950 & 0.0003 & 6234.0 & $-$0.0010 & CCD & I & This paper (SWASP)\\
2454189.6675 & 0.0004 & 6277.0 & $-$0.0015 & CCD & I & This paper (SWASP)\\
2454190.5743 & 0.0003 & 6278.0 & $-$0.0011 & CCD & I & This paper (SWASP)\\
2454200.5415 & 0.002 & 6289.0 & $-$0.0037 & CCD & I & \citet{Hubscher2007IBVS.5802....1H}\\
2454208.7015 & 0.0005 & 6298.0 & $-$0.0009 & CCD & I & This paper (SWASP)\\
2454210.5136 & 0.0003 & 6300.0 & $-$0.0015 & CCD & I & This paper (SWASP)\\
2454215.4993$^{*}$ & 0.0077 & 6305.5 & $-$0.0007 & CCD & II & This paper (SWASP)\\
2454218.6699 & 0.0008 & 6309.0 & $-$0.0023 & CCD & I & This paper (SWASP)\\
2454219.5779 & 0.0003 & 6310.0 & $-$0.0007 & CCD & I & This paper (SWASP)\\
2454220.4840 & 0.001 & 6311.0 & $-$0.0009 & CCD & I & This paper (SWASP)\\
2454224.5636$^{*}$ & 0.0016 & 6315.5 & 0.0001 & CCD & II & This paper (SWASP)\\
2454230.4529 & 0.0005 & 6322.0 & $-$0.0018 & CCD & I & This paper (SWASP)\\
2454249.4862 & 0.0004 & 6343.0 & $-$0.0019 & CCD & I & This paper (SWASP)\\
2454254.4684$^{*}$ & 0.0008 & 6348.5 & $-$0.0046 & CCD & II & This paper (SWASP)\\
2454259.4575 & 0.001 & 6354.0 & $-$0.0004 & CCD & I & \citet{Hubscher2007IBVS.5802....1H}\\
2454268.5203 & 0.0002 & 6364.0 & $-$0.0011 & CCD & I & This paper (SWASP)\\
2454269.4269 & 0.0004 & 6365.0 & $-$0.0009 & CCD & I & This paper (SWASP)\\
2454273.5029$^{*}$ & 0.0012 & 6369.5 & $-$0.0034 & CCD & II & This paper (SWASP)\\
2454278.4908 & 0.0004 & 6375.0 & $-$0.0005 & CCD & I & This paper (SWASP)\\
2454937.4072 & 0.0004 & 7102.0 & 0.0000 & CCD & I & \citet{Brat+2009OEJV..107....1B}\\
2454961.8786 & 0.0003 & 7129.0 & 0.0000 & CCD & I & \citet{Diethelm_2009IBVS.5894....1D}\\
2455073.3614 & 0.0002 & 7252.0 & 0.0018 & PE & I & \citet{Huebscher+2011BAVSM.214....1H}\\
2455314.4506 & 0.0075 & 7518.0 & 0.0021 & PE & I & \citet{Huebscher+2011BAVSM.214....1H}\\
2455642.5507 & 0.0044 & 7880.0 & 0.0037 & CCD & I & \citet{Hubscher+2012IBVS.6010....1H}\\
2455643.4581 & 0.0005 & 7881.0 & 0.0048 & CCD & I & \citet{Honkova+2013OEJV..160....1H}\\
2455648.8966 & 0.0005 & 7887.0 & 0.0052 & CCD & I & \citet{Diethelm_2011IBVS.5992....1D}\\
2455697.8400 & 0.0005 & 7941.0 & 0.0057 & CCD & I & \citet{Diethelm_2011IBVS.5992....1D}\\
2455956.6020$^{*}$ & 0.0008 & 8226.5 & 0.0050 & CCD & II & \citet{Honkova+2013OEJV..160....1H}\\
2456035.9068 & 0.0007 & 8314.0 & 0.0043 & CCD & I & \citet{Diethelm_2012IBVS.6029....1D}\\
2456069.4404 & 0.0035 & 8351.0 & 0.0029 & CCD & I & \citet{Hubscher+2013IBVS.6070....1H}\\
2456089.3815 & 0.0002 & 8373.0 & 0.0043 & CCD & I & \citet{Dogruel+2015NewA...40...20D}\\
2456133.3421$^{*}$ & 0.0002 & 8421.5 & 0.0070 & CCD & II & \citet{Dogruel+2015NewA...40...20D}\\
2456358.5673 & 0.0003 & 8670.0 & 0.0044 & CCD & I & \citet{Dogruel+2015NewA...40...20D}\\
2456408.4088$^{*}$ & 0.0003 & 8725.0 & $-$0.0033 & CCD & I & \citet{Honkova+2013OEJV..160....1H}\\
2456417.4794 & 0.0001 & 8735.0 & 0.0038 & CCD & I & \citet{Dogruel+2015NewA...40...20D}\\
2456730.6227$^{*}$ & 0.0117 & 9080.5 & 0.0034 & CCD & II & \citet{Hubscher+2015IBVS.6149....1H}\\
2456736.5142 & 0.0037 & 9087.0 & 0.0036 & CCD & I & \citet{Hubscher+2015IBVS.6149....1H}\\
2456765.5180 & 0.0006 & 9119.0 & 0.0043 & CCD & I & \citet{Honkova+2015OEJV..168....1H}\\
2456792.2589$^{*}$ & - & 9148.5 & 0.0079 & CCD & II & \citet{Zhang+2015AJ....149...96Z}\\
2456798.1462 & - & 9155.0 & 0.0039 & CCD & I & \citet{Zhang+2015AJ....149...96Z}\\
2456803.1325$^{*}$ & - & 9160.5 & 0.0053 & CCD & II & \citet{Zhang+2015AJ....149...96Z}\\
2456807.2099 & - & 9165.0 & 0.0041 & CCD & I & \citet{Zhang+2015AJ....149...96Z}\\
2457073.6750 & 0.0006 & 9459.0 & 0.0025 & CCD & I & \citet{Jurysek+2017OEJV..179....1J}\\
2457119.4469$^{*}$ & 0.0052 & 9509.5 & 0.0038 & CCD & II & \citet{Hubscher_2017IBVS.6196....1H}\\
2457494.2197 & - & 9923.0 & 0.0011 & CCD & I & \citet{Nagai_2017}\\
2457500.5677 & 0.0048 & 9930.0 & 0.0047 & CCD & I & \citet{Hubscher_2017IBVS.6196....1H}\\
2457838.6286 & 0.0019 & 10303.0 & $-$0.0027 & PE & I & \citet{Pagel_2018IBVS.6244....1P}\\
2457897.5488 & 0.008 & 10368.0 & 0.0048 & CCD & I & \citet{Paschke_2018}\\
2458928.5149$^{*}$ & 0.0007 & 11505.5 & $-$0.0014 & CCD & II & This paper (TESS)\\
2458928.9659 & 0.0001 & 11506.0 & $-$0.0036 & CCD & I & This paper (TESS)\\
2458929.4238$^{*}$ & 0.0005 & 11506.5 & 0.0011 & CCD & II & This paper (TESS)\\
2458929.8724 & 0.0001 & 11507.0 & $-$0.0035 & CCD & I & This paper (TESS)\\
2458930.3261$^{*}$ & 0.0003 & 11507.5 & $-$0.0029 & CCD & II & This paper (TESS)\\
2458930.7785 & 0.0002 & 11508.0 & $-$0.0037 & CCD & I & This paper (TESS)\\
2458931.2325$^{*}$ & 0.0007 & 11508.5 & $-$0.0029 & CCD & II & This paper (TESS)\\
2458931.6851 & 0.0001 & 11509.0 & $-$0.0035 & CCD & I & This paper (TESS)\\
2458932.1378$^{*}$ & 0.0002 & 11509.5 & $-$0.0039 & CCD & II & This paper (TESS)\\
2458932.5914 & 0.0002 & 11510.0 & $-$0.0035 & CCD & I & This paper (TESS)\\
2458933.0438$^{*}$ & 0.0005 & 11510.5 & $-$0.0043 & CCD & II & This paper (TESS)\\
2458933.4979 & 0.0002 & 11511.0 & $-$0.0034 & CCD & I & This paper (TESS)\\
2458933.9581$^{*}$ & 0.0007 & 11511.5 & 0.0037 & CCD & II & This paper (TESS)\\
2458934.4041 & 0.0002 & 11512.0 & $-$0.0035 & CCD & I & This paper (TESS)\\
2458934.8559$^{*}$ & 0.0012 & 11512.5 & $-$0.0049 & CCD & II & This paper (TESS)\\
2458935.3103 & 0.0003 & 11513.0 & $-$0.0037 & CCD & I & This paper (TESS)\\
2458935.7673$^{*}$ & 0.0004 & 11513.5 & 0.0002 & CCD & II & This paper (TESS)\\
2458936.2165 & 0.0003 & 11514.0 & $-$0.0038 & CCD & I & This paper (TESS)\\
2458936.6704$^{*}$ & 0.0001 & 11514.5 & $-$0.0031 & CCD & II & This paper (TESS)\\
2458937.1231 & 0.0004 & 11515.0 & $-$0.0036 & CCD & I & This paper (TESS)\\
2458937.5771$^{*}$ & 0.0005 & 11515.5 & $-$0.0027 & CCD & II & This paper (TESS)\\
2458938.0295 & 0.0002 & 11516.0 & $-$0.0035 & CCD & I & This paper (TESS)\\
2458938.4856$^{*}$ & 0.0005 & 11516.5 & $-$0.0006 & CCD & II & This paper (TESS)\\
2458938.9360 & 0.0003 & 11517.0 & $-$0.0034 & CCD & I & This paper (TESS)\\
2458939.3845$^{*}$ & 0.0008 & 11517.5 & $-$0.0080 & CCD & II & This paper (TESS)\\
2458939.8421 & 0.0001 & 11518.0 & $-$0.0036 & CCD & I & This paper (TESS)\\
2458940.2995$^{*}$ & 0.0006 & 11518.5 & 0.0006 & CCD & II & This paper (TESS)\\
2458940.7484 & 0.0001 & 11519.0 & $-$0.0037 & CCD & I & This paper (TESS)\\
2458942.1107$^{*}$ & 0.0003 & 11520.5 & $-$0.0009 & CCD & II & This paper (TESS)\\
2458942.5616 & 0.0004 & 11521.0 & $-$0.0032 & CCD & I & This paper (TESS)\\
2458943.0151$^{*}$ & 0.0005 & 11521.5 & $-$0.0028 & CCD & II & This paper (TESS)\\
2458943.4679 & 0.0003 & 11522.0 & $-$0.0032 & CCD & I & This paper (TESS)\\
2458943.9184$^{*}$ & 0.0006 & 11522.5 & $-$0.0059 & CCD & II & This paper (TESS)\\
2458944.3742 & 0.0002 & 11523.0 & $-$0.0033 & CCD & I & This paper (TESS)\\
2458944.8281$^{*}$ & 0.0007 & 11523.5 & $-$0.0025 & CCD & II & This paper (TESS)\\
2458945.2804 & 0.0001 & 11524.0 & $-$0.0034 & CCD & I & This paper (TESS)\\
2458945.7260$^{*}$ & 0.001 & 11524.5 & $-$0.0110 & CCD & II & This paper (TESS)\\
2458946.1864 & 0.0001 & 11525.0 & $-$0.0038 & CCD & I & This paper (TESS)\\
2458946.6406$^{*}$ & 0.0004 & 11525.5 & $-$0.0027 & CCD & II & This paper (TESS)\\
2458947.0930 & 0.0003 & 11526.0 & $-$0.0035 & CCD & I & This paper (TESS)\\
2458947.5549$^{*}$ & 0.001 & 11526.5 & 0.0052 & CCD & II & This paper (TESS)\\
2458947.9993 & 0.0002 & 11527.0 & $-$0.0035 & CCD & I & This paper (TESS)\\
2458948.4523$^{*}$ & 0.0001 & 11527.5 & $-$0.0037 & CCD & II & This paper (TESS)\\
2458948.9058 & 0.0002 & 11528.0 & $-$0.0034 & CCD & I & This paper (TESS)\\
2458949.3580$^{*}$ & 0.0004 & 11528.5 & $-$0.0044 & CCD & II & This paper (TESS)\\
2458949.8120 & 0.0002 & 11529.0 & $-$0.0035 & CCD & I & This paper (TESS)\\
2458950.2674$^{*}$ & 0.0005 & 11529.5 & $-$0.0013 & CCD & II & This paper (TESS)\\
2458950.7182 & 0.0001 & 11530.0 & $-$0.0037 & CCD & I & This paper (TESS)\\
2458951.1703$^{*}$ & 0.0007 & 11530.5 & $-$0.0048 & CCD & II & This paper (TESS)\\
2458951.6242 & 0.0002 & 11531.0 & $-$0.0040 & CCD & I & This paper (TESS)\\
2458952.0877$^{*}$ & 0.0013 & 11531.5 & 0.0063 & CCD & II & This paper (TESS)\\
2458952.5304 & 0.0003 & 11532.0 & $-$0.0042 & CCD & I & This paper (TESS)\\
2458952.9846$^{*}$ & 0.0008 & 11532.5 & $-$0.0032 & CCD & II & This paper (TESS)\\
2458953.4374 & 0.0003 & 11533.0 & $-$0.0035 & CCD & I & This paper (TESS)\\
2458953.8959$^{*}$ & 0.0009 & 11533.5 & 0.0018 & CCD & II & This paper (TESS)\\
2458954.3437 & 0.0002 & 11534.0 & $-$0.0036 & CCD & I & This paper (TESS)\\
2458954.7930$^{*}$ & 0.0005 & 11534.5 & $-$0.0075 & CCD & II & This paper (TESS)\\
2458956.1565 & 0.0001 & 11536.0 & $-$0.0035 & CCD & I & This paper (TESS)\\
2458956.6117$^{*}$ & 0.0004 & 11536.5 & $-$0.0015 & CCD & II & This paper (TESS)\\
2458957.0627 & 0.0002 & 11537.0 & $-$0.0036 & CCD & I & This paper (TESS)\\
2458957.5202$^{*}$ & 0.0007 & 11537.5 & 0.0007 & CCD & II & This paper (TESS)\\
2458957.9692 & 0.0003 & 11538.0 & $-$0.0035 & CCD & I & This paper (TESS)\\
2458958.4215$^{*}$ & 0.0006 & 11538.5 & $-$0.0044 & CCD & II & This paper (TESS)\\
2458958.8751 & 0.0003 & 11539.0 & $-$0.0039 & CCD & I & This paper (TESS)\\
2458959.3285$^{*}$ & 0.0004 & 11539.5 & $-$0.0037 & CCD & II & This paper (TESS)\\
2458959.7822 & 0.0003 & 11540.0 & $-$0.0032 & CCD & I & This paper (TESS)\\
2458960.2324$^{*}$ & 0.0004 & 11540.5 & $-$0.0062 & CCD & II & This paper (TESS)\\
2458960.6884 & 0.0002 & 11541.0 & $-$0.0033 & CCD & I & This paper (TESS)\\
2458961.1357$^{*}$ & 0.0008 & 11541.5 & $-$0.0092 & CCD & II & This paper (TESS)\\
2458961.5943 & 0.0002 & 11542.0 & $-$0.0038 & CCD & I & This paper (TESS)\\
2458962.0498$^{*}$ & 0.0003 & 11542.5 & $-$0.0015 & CCD & II & This paper (TESS)\\
2458962.5007 & 0.0001 & 11543.0 & $-$0.0037 & CCD & I & This paper (TESS)\\
2458962.9459$^{*}$ & 0.0011 & 11543.5 & $-$0.0117 & CCD & II & This paper (TESS)\\
2458963.4073 & 0.0002 & 11544.0 & $-$0.0035 & CCD & I & This paper (TESS)\\
2458963.8609$^{*}$ & 0.0003 & 11544.5 & $-$0.0031 & CCD & II & This paper (TESS)\\
2458964.3136 & 0.0002 & 11545.0 & $-$0.0035 & CCD & I & This paper (TESS)\\
2458964.7658$^{*}$ & 0.0009 & 11545.5 & $-$0.0045 & CCD & II & This paper (TESS)\\
2458965.2198 & 0.0002 & 11546.0 & $-$0.0037 & CCD & I & This paper (TESS)\\
2458965.6749$^{*}$ & 0.0006 & 11546.5 & $-$0.0018 & CCD & II & This paper (TESS)\\
2458966.1263 & 0.0001 & 11547.0 & $-$0.0035 & CCD & I & This paper (TESS)\\
2458966.5763$^{*}$ & 0.0005 & 11547.5 & $-$0.0067 & CCD & II & This paper (TESS)\\
2458967.0323 & 0.0001 & 11548.0 & $-$0.0039 & CCD & I & This paper (TESS)\\
2458967.4809$^{*}$ & 0.0014 & 11548.5 & $-$0.0085 & CCD & II & This paper (TESS)\\
2458967.9386 & 0.0002 & 11549.0 & $-$0.0039 & CCD & I & This paper (TESS)\\
2458969.2871$^{*}$ & 0.0002 & 11550.5 & $-$0.0150 & CCD & II & This paper (TESS)\\
2458969.7514 & 0.0002 & 11551.0 & $-$0.0038 & CCD & I & This paper (TESS)\\
2458970.2022$^{*}$ & 0.0003 & 11551.5 & $-$0.0062 & CCD & II & This paper (TESS)\\
2458970.6581 & 0.0002 & 11552.0 & $-$0.0035 & CCD & I & This paper (TESS)\\
2458971.1163$^{*}$ & 0.0006 & 11552.5 & 0.0015 & CCD & II & This paper (TESS)\\
2458971.5639 & 0.0001 & 11553.0 & $-$0.0040 & CCD & I & This paper (TESS)\\
2458972.0139$^{*}$ & 0.0009 & 11553.5 & $-$0.0072 & CCD & II & This paper (TESS)\\
2458972.4704 & 0.0003 & 11554.0 & $-$0.0039 & CCD & I & This paper (TESS)\\
2458972.9242$^{*}$ & 0.0009 & 11554.5 & $-$0.0033 & CCD & II & This paper (TESS)\\
2458973.3770 & 0.0004 & 11555.0 & $-$0.0036 & CCD & I & This paper (TESS)\\
2458973.8313$^{*}$ & 0.0003 & 11555.5 & $-$0.0025 & CCD & II & This paper (TESS)\\
2458974.2831 & 0.0003 & 11556.0 & $-$0.0039 & CCD & I & This paper (TESS)\\
2458974.7370$^{*}$ & 0.0005 & 11556.5 & $-$0.0032 & CCD & II & This paper (TESS)\\
2458975.1897 & 0.0005 & 11557.0 & $-$0.0036 & CCD & I & This paper (TESS)\\
2458975.6413$^{*}$ & 0.0003 & 11557.5 & $-$0.0052 & CCD & II & This paper (TESS)\\
2458976.0960 & 0.0003 & 11558.0 & $-$0.0037 & CCD & I & This paper (TESS)\\
2458976.5453$^{*}$ & 0.0007 & 11558.5 & $-$0.0076 & CCD & II & This paper (TESS)\\
2458977.0024 & 0.0001 & 11559.0 & $-$0.0036 & CCD & I & This paper (TESS)\\
2458977.4588$^{*}$ & 0.0005 & 11559.5 & $-$0.0004 & CCD & II & This paper (TESS)\\
2458977.9085 & 0.0002 & 11560.0 & $-$0.0039 & CCD & I & This paper (TESS)\\
2458978.3553$^{*}$ & 0.0006 & 11560.5 & $-$0.0103 & CCD & II & This paper (TESS)\\
2458978.8149 & 0.0003 & 11561.0 & $-$0.0038 & CCD & I & This paper (TESS)\\
2458979.2698$^{*}$ & 0.0002 & 11561.5 & $-$0.0021 & CCD & II & This paper (TESS)\\
2458979.7216 & 0.0003 & 11562.0 & $-$0.0035 & CCD & I & This paper (TESS)\\
2458980.1744$^{*}$ & 0.0004 & 11562.5 & $-$0.0038 & CCD & II & This paper (TESS)\\
2458980.6278 & 0.0002 & 11563.0 & $-$0.0036 & CCD & I & This paper (TESS)\\
2458981.0820$^{*}$ & 0.0009 & 11563.5 & $-$0.0026 & CCD & II & This paper (TESS)\\
2458981.5342 & 0.0001 & 11564.0 & $-$0.0036 & CCD & I & This paper (TESS)\\
2458981.9918$^{*}$ & 0.0011 & 11564.5 & 0.0009 & CCD & II & This paper (TESS)
\end{longtable}
\tablecomments{\\
$^{*}$ The timings were not used in our timing analysis \\
$^{a}$ CCD: Electronic camera, PE: Photoelectric \\
$^{b}$ I: Primary eclipse, II: Secondary eclipse}

\newpage

\section{Lists of multiple frequencies} \label{sec:appendixB}
\setcounter{table}{0} \renewcommand{\thetable}{B.\arabic{table}}

\begin{longtable}{lcccccc}
\tablewidth{0pt}
\caption{The results of the multiple frequency analysis of EW Boo} \label{tab:freq_all}\\
\hline\hline
\colhead{} & \colhead{Frequency (d$^{-1}$)} &  \colhead{Amplitude (mmag)} &  \colhead{Phase (rad)} &  \colhead{S/N} &  \colhead{Combination}\\
\hline
\endfirsthead
\caption{continued.}\\
\hline\hline
\colhead{} & \colhead{Frequency (d$^{-1}$)} &  \colhead{Amplitude (mmag)} &  \colhead{Phase (rad)} &  \colhead{S/N} & \colhead{Combination}\\
\hline
\endhead
\hline
\endfoot
$f_{1}$ & 52.37033 $ \pm $ 0.00005 & 4.20 $ \pm $ 0.04 & 1.51 $ \pm $ 0.03 & 193.03 & \\
$f_{2}$ & 49.19283 $ \pm $ 0.00007 & 3.31 $ \pm $ 0.04 & 1.70 $ \pm $ 0.04 & 141.36 & \\
$f_{3}$ & 46.98566 $ \pm $ 0.00007 & 3.19 $ \pm $ 0.04 & 5.63 $ \pm $ 0.04 & 131.21 & $f_{2}$ $-$ 2$f_{\rm orb}$ \\
$f_{4}$ & 46.33846 $ \pm $ 0.00012 & 2.01 $ \pm $ 0.04 & 4.06 $ \pm $ 0.06 & 80.31 & \\
$f_{5}$ & 40.99743 $ \pm $ 0.00011 & 1.88 $ \pm $ 0.04 & 2.63 $ \pm $ 0.06 & 87.41 & \\
$f_{6}$ & 43.20367 $ \pm $ 0.00015 & 1.55 $ \pm $ 0.04 & 0.11 $ \pm $ 0.08 & 64.07 & $f_{5}$ $+$ 2$f_{\rm orb}$ \\
$f_{7}$ & 41.27599 $ \pm $ 0.00018 & 1.13 $ \pm $ 0.04 & 5.70 $ \pm $ 0.09 & 51.69 & \\
$f_{8}$ & 49.30983 $ \pm $ 0.00023 & 0.96 $ \pm $ 0.04 & 5.01 $ \pm $ 0.12 & 41.23 & \\
$f_{9}$ & 43.48223 $ \pm $ 0.00024 & 0.95 $ \pm $ 0.04 & 0.11 $ \pm $ 0.13 & 39.23 & \\
$f_{10}$ & 47.28187 $ \pm $ 0.00035 & 0.65 $ \pm $ 0.04 & 1.95 $ \pm $ 0.18 & 27.08 & \\
$f_{11}$ & 4.41341 $ \pm $ 0.00029 & 0.86 $ \pm $ 0.04 & 4.59 $ \pm $ 0.15 & 32.72 & 4$f_{\rm orb}$ \\
$f_{12}$ & 5.51653 $ \pm $ 0.00029 & 0.80 $ \pm $ 0.04 & 3.18 $ \pm $ 0.15 & 33.15 & 5$f_{\rm orb}$ \\
$f_{13}$ & 44.27615 $ \pm $ 0.00041 & 0.56 $ \pm $ 0.04 & 0.37 $ \pm $ 0.22 & 23.04 & \\
$f_{14}$ & 40.15337 $ \pm $ 0.00034 & 0.59 $ \pm $ 0.04 & 2.76 $ \pm $ 0.18 & 27.80 & \\
$f_{15}$ & 49.48904 $ \pm $ 0.00035 & 0.63 $ \pm $ 0.04 & 1.17 $ \pm $ 0.18 & 27.31 & \\
$f_{16}$ & 3.31122 $ \pm $ 0.00052 & 0.49 $ \pm $ 0.05 & 3.73 $ \pm $ 0.28 & 18.13 & 3$f_{\rm orb}$ \\
$f_{17}$ & 40.52944 $ \pm $ 0.00031 & 0.67 $ \pm $ 0.04 & 5.59 $ \pm $ 0.16 & 30.81 & \\
$f_{18}$ & 38.32320 $ \pm $ 0.00032 & 0.64 $ \pm $ 0.04 & 1.68 $ \pm $ 0.17 & 29.40 & \\
$f_{19}$ & 46.48238 $ \pm $ 0.00049 & 0.49 $ \pm $ 0.04 & 1.28 $ \pm $ 0.26 & 19.48 & \\
$f_{20}$ & 42.35961 $ \pm $ 0.00052 & 0.43 $ \pm $ 0.04 & 3.52 $ \pm $ 0.28 & 18.17 & \\
$f_{21}$ & 2.19974 $ \pm $ 0.00072 & 0.37 $ \pm $ 0.05 & 2.47 $ \pm $ 0.38 & 13.15 & $f_{2}$ $-$ $f_{3}$ \\
$f_{22}$ & 44.90385 $ \pm $ 0.00060 & 0.39 $ \pm $ 0.04 & 5.62 $ \pm $ 0.32 & 15.78 & $f_{8}$ $-$ $f_{11}$ \\
$f_{23}$ & 47.89843 $ \pm $ 0.00049 & 0.46 $ \pm $ 0.04 & 2.32 $ \pm $ 0.26 & 19.24 & $f_{11}$ $+$ $f_{9}$ \\
$f_{24}$ & 40.33073 $ \pm $ 0.00059 & 0.35 $ \pm $ 0.04 & 5.16 $ \pm $ 0.31 & 16.07 & \\
$f_{25}$ & 45.69126 $ \pm $ 0.00059 & 0.41 $ \pm $ 0.04 & 0.15 $ \pm $ 0.31 & 16.15 & $f_{11}$ $+$ $f_{7}$ \\
$f_{26}$ & 44.16750 $ \pm $ 0.00060 & 0.38 $ \pm $ 0.04 & 5.88 $ \pm $ 0.32 & 15.80 & \\
$f_{27}$ & 1.11241 $ \pm $ 0.00057 & 0.48 $ \pm $ 0.05 & 1.67 $ \pm $ 0.30 & 16.58 & $f_{12}$ $-$ $f_{11}$ \\
$f_{28}$ & 1.15419 $ \pm $ 0.00068 & 0.41 $ \pm $ 0.05 & 3.70 $ \pm $ 0.36 & 14.06 & \\
$f_{29}$ & 1.05298 $ \pm $ 0.00101 & 0.27 $ \pm $ 0.05 & 1.92 $ \pm $ 0.53 & 9.45 & $f_{13}$ $-$ $f_{6}$ \\
$f_{30}$ & 1.21733 $ \pm $ 0.00066 & 0.41 $ \pm $ 0.05 & 1.48 $ \pm $ 0.35 & 14.33 & \\
$f_{31}$ & 0.98705 $ \pm $ 0.00100 & 0.27 $ \pm $ 0.05 & 3.24 $ \pm $ 0.53 & 9.50 & $f_{26}$ $-$ $f_{6}$ \\
$f_{32}$ & 39.04190 $ \pm $ 0.00081 & 0.26 $ \pm $ 0.04 & 5.91 $ \pm $ 0.43 & 11.66 & $f_{9}$ $-$ $f_{11}$ \\
$f_{33}$ & 36.11510 $ \pm $ 0.00082 & 0.26 $ \pm $ 0.04 & 1.04 $ \pm $ 0.43 & 11.64 & $f_{17}$ $-$ $f_{11}$ \\
$f_{34}$ & 9.92994 $ \pm $ 0.00065 & 0.34 $ \pm $ 0.04 & 6.13 $ \pm $ 0.35 & 14.55 & $f_{11}$ $+$ $f_{12}$ \\
$f_{35}$ & 46.22146 $ \pm $ 0.00091 & 0.26 $ \pm $ 0.04 & 4.26 $ \pm $ 0.48 & 10.45 & $f_{10}$ $-$ $f_{29}$ \\
$f_{36}$ & 44.01337 $ \pm $ 0.00096 & 0.24 $ \pm $ 0.04 & 3.62 $ \pm $ 0.51 & 9.85 & $f_{35}$ $-$ $f_{21}$ \\
$f_{37}$ & 1.09105 $ \pm $ 0.00098 & 0.28 $ \pm $ 0.05 & 4.10 $ \pm $ 0.52 & 9.69 & $f_{27}$ \\
$f_{38}$ & 0.64627 $ \pm $ 0.00158 & 0.17 $ \pm $ 0.05 & 2.44 $ \pm $ 0.83 & 6.02 & $f_{3}$ $-$ $f_{4}$ \\
$f_{39}$ & 37.08358 $ \pm $ 0.00105 & 0.20 $ \pm $ 0.04 & 1.25 $ \pm $ 0.55 & 9.09 & $f_{3}$ $-$ $f_{34}$ \\
$f_{40}$ & 48.08878 $ \pm $ 0.00100 & 0.22 $ \pm $ 0.04 & 4.41 $ \pm $ 0.53 & 9.51 & $f_{2}$ $-$ $f_{27}$ \\
$f_{41}$ & 0.17085 $ \pm $ 0.00163 & 0.16 $ \pm $ 0.05 & 1.95 $ \pm $ 0.86 & 5.81 & $f_{19}$ $-$ $f_{4}$\\
$f_{42}$ & 104.74059 $ \pm $ 0.00083 & 0.18 $ \pm $ 0.03 & 3.93 $ \pm $ 0.44 & 11.38 & 2$f_{1}$ \\
$f_{43}$ & 8.82774 $ \pm $ 0.00085 & 0.26 $ \pm $ 0.04 & 5.65 $ \pm $ 0.45 & 11.23 & 8$f_{\rm orb}$ \\
$f_{44}$ & 1.02698 $ \pm $ 0.00113 & 0.24 $ \pm $ 0.05 & 3.17 $ \pm $ 0.60 & 8.44 & $f_{29}$ \\
$f_{45}$ & 1.17276 $ \pm $ 0.00117 & 0.24 $ \pm $ 0.05 & 3.33 $ \pm $ 0.62 & 8.13 & $f_{28}$\\
$f_{46}$ & 50.73701 $ \pm $ 0.00133 & 0.17 $ \pm $ 0.04 & 0.68 $ \pm $ 0.70 & 7.17 & $f_{11}$ $+$ $f_{4}$ \\
$f_{47}$ & 36.26738 $ \pm $ 0.00108 & 0.19 $ \pm $ 0.04 & 4.42 $ \pm $ 0.57 & 8.79 & $f_{35}$ $-$ $f_{34}$ \\
$f_{48}$ & 46.12211 $ \pm $ 0.00177 & 0.14 $ \pm $ 0.04 & 0.52 $ \pm $ 0.93 & 5.37 & $f_{10}$ $-$ $f_{28}$ \\
$f_{49}$ & 0.14207 $ \pm $ 0.00114 & 0.23 $ \pm $ 0.05 & 1.43 $ \pm $ 0.60 & 8.37 & $f_{8}$ $-$ $f_{2}$ \\
$f_{50}$ & 0.19314 $ \pm $ 0.00185 & 0.14 $ \pm $ 0.05 & 0.77 $ \pm $ 0.97 & 5.14 & $f_{41}$ \\
$f_{51}$ & 44.94192 $ \pm $ 0.00158 & 0.15 $ \pm $ 0.04 & 4.10 $ \pm $ 0.84 & 6.00 & $f_{11}$ $+$ $f_{17}$ \\
$f_{52}$ & 46.37282 $ \pm $ 0.00133 & 0.18 $ \pm $ 0.04 & 5.17 $ \pm $ 0.70 & 7.12 & $f_{21}$ $+$ $f_{26}$ \\
$f_{53}$ & 41.96498 $ \pm $ 0.00130 & 0.17 $ \pm $ 0.04 & 1.60 $ \pm $ 0.68 & 7.32 & $f_{31}$ $+$ $f_{5}$ \\
$f_{54}$ & 12.13803 $ \pm $ 0.00109 & 0.21 $ \pm $ 0.04 & 0.29 $ \pm $ 0.58 & 8.72 & 11$f_{\rm orb}$ \\
$f_{55}$ & 98.71150 $ \pm $ 0.00114 & 0.14 $ \pm $ 0.03 & 1.50 $ \pm $ 0.60 & 8.32 & $f_{1}$ $+$ $f_{4}$ \\
$f_{56}$ & 38.95090 $ \pm $ 0.00145 & 0.14 $ \pm $ 0.04 & 1.29 $ \pm $ 0.77 & 6.54 & $f_{14}$ $-$ $f_{30}$ \\
$f_{57}$ & 99.35592 $ \pm $ 0.00098 & 0.16 $ \pm $ 0.03 & 1.68 $ \pm $ 0.52 & 9.65 & $f_{1}$ $+$ $f_{3}$ \\
$f_{58}$ & 101.56402 $ \pm $ 0.00104 & 0.16 $ \pm $ 0.03 & 2.05 $ \pm $ 0.55 & 9.09 & $f_{1}$ $+$ $f_{2}$ \\
$f_{59}$ & 0.12535 $ \pm $ 0.00183 & 0.14 $ \pm $ 0.05 & 4.40 $ \pm $ 0.97 & 5.19 & $f_{49}$ \\
$f_{60}$ & 0.57292 $ \pm $ 0.00206 & 0.13 $ \pm $ 0.05 & 3.30 $ \pm $ 1.09 & 4.62 & $f_{30}$ $-$ $f_{38}$ \\
$f_{61}$ & 45.23534 $ \pm $ 0.00168 & 0.14 $ \pm $ 0.04 & 2.49 $ \pm $ 0.89 & 5.64 & $f_{4}$ $-$ $f_{27}$ \\
$f_{62}$ & 41.99005 $ \pm $ 0.00140 & 0.16 $ \pm $ 0.04 & 1.46 $ \pm $ 0.74 & 6.78 & $f_{53}$\\
$f_{63}$ & 39.16447 $ \pm $ 0.00170 & 0.12 $ \pm $ 0.04 & 4.84 $ \pm $ 0.90 & 5.59 & $f_{14}$ $-$ $f_{31}$ \\
$f_{64}$ & 0.62492 $ \pm $ 0.00132 & 0.21 $ \pm $ 0.05 & 1.79 $ \pm $ 0.70 & 7.21 & $f_{38}$ \\
$f_{65}$ & 0.92669 $ \pm $ 0.00152 & 0.18 $ \pm $ 0.05 & 0.97 $ \pm $ 0.80 & 6.25 & $f_{23}$ $-$ $f_{3}$ \\
$f_{66}$ & 0.69827 $ \pm $ 0.00149 & 0.18 $ \pm $ 0.05 & 0.72 $ \pm $ 0.79 & 6.38 & $f_{26}$ $-$ $f_{9}$ \\
$f_{67}$ & 1.07155 $ \pm $ 0.00156 & 0.18 $ \pm $ 0.05 & 3.09 $ \pm $ 0.82 & 6.10 & $f_{29}$\\
$f_{68}$ & 40.47187 $ \pm $ 0.00160 & 0.13 $ \pm $ 0.04 & 1.66 $ \pm $ 0.84 & 5.95 & $f_{22}$ $-$ $f_{11}$ \\
$f_{69}$ & 38.99361 $ \pm $ 0.00169 & 0.13 $ \pm $ 0.04 & 0.81 $ \pm $ 0.89 & 5.62 & $f_{14}$ $-$ $f_{28}$ \\
$f_{70}$ & 1.29162 $ \pm $ 0.00185 & 0.15 $ \pm $ 0.05 & 5.35 $ \pm $ 0.97 & 5.13 & 2$f_{38}$ \\
$f_{71}$ & 0.85055 $ \pm $ 0.00235 & 0.12 $ \pm $ 0.05 & 1.15 $ \pm $ 1.24 & 4.05 & $f_{5}$ $-$ $f_{14}$ \\
$f_{72}$ & 0.77163 $ \pm $ 0.00177 & 0.15 $ \pm $ 0.05 & 5.97 $ \pm $ 0.94 & 5.36 & $f_{3}$ $-$ $f_{35}$ \\
$f_{73}$ & 0.55249 $ \pm $ 0.00226 & 0.12 $ \pm $ 0.05 & 1.25 $ \pm $ 1.19 & 4.20 & $f_{60}$ \\
$f_{74}$ & 16.55051 $ \pm $ 0.00116 & 0.15 $ \pm $ 0.03 & 5.20 $ \pm $ 0.62 & 8.18 & $f_{11}$ $+$ $f_{54}$ \\
$f_{75}$ & 18.75768 $ \pm $ 0.00121 & 0.15 $ \pm $ 0.03 & 1.21 $ \pm $ 0.64 & 7.85 & $f_{34}$ $+$ $f_{43}$ \\
$f_{76}$ & 41.04014 $ \pm $ 0.00137 & 0.15 $ \pm $ 0.04 & 3.88 $ \pm $ 0.72 & 6.94 & $f_{24}$ $+$ $f_{66}$ \\
$f_{77}$ & 40.82379 $ \pm $ 0.00155 & 0.13 $ \pm $ 0.04 & 0.67 $ \pm $ 0.82 & 6.13 & $f_{4}$ $-$ $f_{12}$ \\
$f_{78}$ & 90.18833 $ \pm $ 0.00144 & 0.12 $ \pm $ 0.03 & 2.00 $ \pm $ 0.76 & 6.58 & $f_{2}$ $+$ $f_{5}$ \\
$f_{79}$ & 42.09962 $ \pm $ 0.00155 & 0.14 $ \pm $ 0.04 & 2.98 $ \pm $ 0.82 & 6.13 & $f_{27}$ $+$ $f_{5}$ \\
$f_{80}$ & 2.94444 $ \pm $ 0.00268 & 0.10 $ \pm $ 0.05 & 2.48 $ \pm $ 1.41 & 3.54 & $f_{8}$ $-$ $f_{4}$ \\
$f_{81}$ & 44.88435 $ \pm $ 0.00204 & 0.12 $ \pm $ 0.04 & 5.48 $ \pm $ 1.07 & 4.67 & $f_{22}$ \\
$f_{82}$ & 0.09286 $ \pm $ 0.00181 & 0.15 $ \pm $ 0.05 & 5.69 $ \pm $ 0.96 & 5.24 & $f_{8}$ $-$ $f_{2}$ \\
$f_{83}$ & 39.78381 $ \pm $ 0.00188 & 0.11 $ \pm $ 0.04 & 4.16 $ \pm $ 0.99 & 5.06 & $f_{5}$ $-$ $f_{30}$ \\
$f_{84}$ & 0.02229 $ \pm $ 0.00173 & 0.15 $ \pm $ 0.05 & 3.10 $ \pm $ 0.92 & 5.48 & $f_{52}$ $-$ $f_{4}$ \\
$f_{85}$ & 2.35388 $ \pm $ 0.00206 & 0.13 $ \pm $ 0.05 & 1.75 $ \pm $ 1.09 & 4.60 & 2$f_{45}$ \\
$f_{86}$ & 1.36404 $ \pm $ 0.00228 & 0.12 $ \pm $ 0.05 & 6.23 $ \pm $ 1.21 & 4.16 & $f_{20}$ $-$ $f_{5}$ \\
$f_{87}$ & 0.87377 $ \pm $ 0.00208 & 0.13 $ \pm $ 0.05 & 3.86 $ \pm $ 1.10 & 4.56 & $f_{71}$\\
$f_{88}$ & 1.32504 $ \pm $ 0.00211 & 0.13 $ \pm $ 0.05 & 4.33 $ \pm $ 1.11 & 4.51 & $f_{35}$ $-$ $f_{22}$ \\
$f_{89}$ & 96.66033 $ \pm $ 0.00156 & 0.11 $ \pm $ 0.03 & 5.01 $ \pm $ 0.82 & 6.08 & $f_{1}$ $+$ $f_{13}$ \\
$f_{90}$ & 40.23602 $ \pm $ 0.00185 & 0.11 $ \pm $ 0.04 & 3.59 $ \pm $ 0.98 & 5.13 & $f_{1}$ $-$ $f_{54}$ \\
$f_{91}$ & 0.39928 $ \pm $ 0.00249 & 0.11 $ \pm $ 0.05 & 2.53 $ \pm $ 1.31 & 3.81 & 2$f_{50}$ \\
$f_{92}$ & 34.05929 $ \pm $ 0.00176 & 0.11 $ \pm $ 0.03 & 3.75 $ \pm $ 0.93 & 5.39 & $f_{36}$ $-$ $f_{34}$ \\
$f_{93}$ & 94.79580 $ \pm $ 0.00158 & 0.10 $ \pm $ 0.03 & 3.27 $ \pm $ 0.84 & 6.02 & $f_{42}$ $-$ $f_{34}$ \\
$f_{94}$ & 46.64024 $ \pm $ 0.00249 & 0.10 $ \pm $ 0.04 & 3.74 $ \pm $ 1.31 & 3.81 & $f_{10}$ $-$ $f_{64}$ \\
$f_{95}$ & 39.33718 $ \pm $ 0.00208 & 0.10 $ \pm $ 0.04 & 0.51 $ \pm $ 1.09 & 4.57 & $f_{17}$ $-$ $f_{30}$\\
$f_{96}$ & 41.35956 $ \pm $ 0.00202 & 0.10 $ \pm $ 0.04 & 5.62 $ \pm $ 1.07 & 4.70 & $f_{14}$ $+$ $f_{30}$ \\
$f_{97}$ & 96.18027 $ \pm $ 0.00171 & 0.10 $ \pm $ 0.03 & 4.95 $ \pm $ 0.90 & 5.56 & $f_{2}$ $+$ $f_{3}$ \\
$f_{98}$ & 1.38911 $ \pm $ 0.00291 & 0.09 $ \pm $ 0.05 & 0.76 $ \pm $ 1.53 & 3.27 & $f_{86}$\\
$f_{99}$ & 2.26752 $ \pm $ 0.00262 & 0.10 $ \pm $ 0.05 & 4.09 $ \pm $ 1.38 & 3.63 & $f_{27}$ $+$ $f_{28}$ \\
$f_{100}$ & 101.67823 $ \pm $ 0.00170 & 0.10 $ \pm $ 0.03 & 4.18 $ \pm $ 0.90 & 5.58 & $f_{1}$ $+$ $f_{8}$ \\
$f_{101}$ & 1.63611 $ \pm $ 0.00295 & 0.09 $ \pm $ 0.05 & 4.81 $ \pm $ 1.56 & 3.22 & $f_{1}$ $-$ $f_{46}$ \\
$f_{102}$ & 35.02591 $ \pm $ 0.00190 & 0.11 $ \pm $ 0.04 & 1.36 $ \pm $ 1.00 & 5.01 & $f_{17}$ $-$ $f_{12}$ \\
$f_{103}$ & 42.22312 $ \pm $ 0.00235 & 0.09 $ \pm $ 0.04 & 1.04 $ \pm $ 1.24 & 4.05 & $f_{30}$ $+$ $f_{5}$ \\
$f_{104}$ & 46.14903 $ \pm $ 0.00240 & 0.10 $ \pm $ 0.04 & 4.45 $ \pm $ 1.26 & 3.97 & $f_{48}$\\
$f_{105}$ & 1.85803 $ \pm $ 0.00295 & 0.09 $ \pm $ 0.05 & 0.13 $ \pm $ 1.56 & 3.22 & 2$f_{65}$ \\
$f_{106}$ & 0.21264 $ \pm $ 0.00232 & 0.11 $ \pm $ 0.05 & 4.96 $ \pm $ 1.23 & 4.09 & $f_{50}$\\
$f_{107}$ & 0.25535 $ \pm $ 0.00186 & 0.14 $ \pm $ 0.05 & 6.17 $ \pm $ 0.98 & 5.12 & $f_{7}$ $-$ $f_{5}$ \\
$f_{108}$ & 11.03120 $ \pm $ 0.00177 & 0.13 $ \pm $ 0.04 & 1.46 $ \pm $ 0.94 & 5.37 & 2$f_{12}$ \\
$f_{109}$ & 0.32035 $ \pm $ 0.00226 & 0.12 $ \pm $ 0.05 & 1.46 $ \pm $ 1.19 & 4.20 & 2$f_{41}$ \\
$f_{110}$ & 2.98808 $ \pm $ 0.00275 & 0.10 $ \pm $ 0.05 & 5.09 $ \pm $ 1.45 & 3.46 & $f_{8}$ $-$ $f_{4}$ \\
$f_{111}$ & 1.81996 $ \pm $ 0.00271 & 0.10 $ \pm $ 0.05 & 2.16 $ \pm $ 1.43 & 3.51 & $f_{14}$ $-$ $f_{18}$ \\
$f_{112}$ & 31.42591 $ \pm $ 0.00216 & 0.09 $ \pm $ 0.03 & 5.30 $ \pm $ 1.14 & 4.40 & $f_{96}$ $-$ $f_{34}$ \\
$f_{113}$ & 47.38772 $ \pm $ 0.00253 & 0.09 $ \pm $ 0.04 & 0.53 $ \pm $ 1.34 & 3.75 & $f_{29}$ $+$ $f_{4}$ \\
$f_{114}$ & 46.97266 $ \pm $ 0.00224 & 0.10 $ \pm $ 0.04 & 1.14 $ \pm $ 1.18 & 4.23 & $f_{3}$ \\
$f_{115}$ & 48.32278 $ \pm $ 0.00249 & 0.09 $ \pm $ 0.04 & 6.06 $ \pm $ 1.31 & 3.81 & $f_{2}$ $-$ $f_{71}$ \\
$f_{116}$ & 41.01136 $ \pm $ 0.00197 & 0.10 $ \pm $ 0.04 & 5.00 $ \pm $ 1.04 & 4.83 & $f_{5}$\\
$f_{117}$ & 1.94346 $ \pm $ 0.00252 & 0.11 $ \pm $ 0.05 & 2.29 $ \pm $ 1.33 & 3.77 & $f_{6}$ $-$ $f_{7}$ \\
$f_{118}$ & 45.62626 $ \pm $ 0.00810 & 0.03 $ \pm $ 0.04 & 3.75 $ \pm $ 4.28 & 1.17 & $f_{3}$ $-$ $f_{86}$ \\
$f_{119}$ & 45.62626 $ \pm $ 0.00362 & 0.07 $ \pm $ 0.04 & 4.54 $ \pm $ 1.91 & 2.62 & $f_{118}$ \\
$f_{120}$ & 0.42249 $ \pm $ 0.00300 & 0.09 $ \pm $ 0.05 & 5.96 $ \pm $ 1.58 & 3.16 & $f_{91}$\\
$f_{121}$ & 37.23958 $ \pm $ 0.00238 & 0.09 $ \pm $ 0.04 & 4.09 $ \pm $ 1.26 & 4.00 & $f_{17}$ $-$ $f_{16}$\\
$f_{122}$ & 33.42880 $ \pm $ 0.00223 & 0.08 $ \pm $ 0.03 & 5.62 $ \pm $ 1.18 & 4.26 & $f_{56}$ $-$ $f_{12}$ \\
$f_{123}$ & 46.32453 $ \pm $ 0.00260 & 0.09 $ \pm $ 0.04 & 1.65 $ \pm $ 1.37 & 3.66 & $f_{4}$ \\
$f_{124}$ & 1.26005 $ \pm $ 0.00276 & 0.10 $ \pm $ 0.05 & 3.66 $ \pm $ 1.45 & 3.45 & 2$f_{64}$ \\
$f_{125}$ & 0.04921 $ \pm $ 0.00253 & 0.11 $ \pm $ 0.05 & 5.27 $ \pm $ 1.34 & 3.75 & $f_{84}$\\
$f_{126}$ & 2.08553 $ \pm $ 0.00290 & 0.09 $ \pm $ 0.05 & 2.33 $ \pm $ 1.53 & 3.28 & 2$f_{29}$ \\
$f_{127}$ & 55.68341 $ \pm $ 0.00227 & 0.08 $ \pm $ 0.03 & 5.79 $ \pm $ 1.19 & 4.19 & $f_{1}$ $+$ $f_{16}$ \\
\end{longtable}

\end{appendix}

\end{document}